# Covalent Grafting of Polyoxometalate Hybrids onto Flat Silicon/Silicon Oxide: Insights from POMs Layers on Oxides

*Maxime Laurans,[a] Kelly Trinh,[a,d] Kevin Dalla Francesca,[a] Guillaume Izzet,[a] Sandra Alves,[a] Etienne Derat,[a] Vincent Humblot,[c,†] Olivier Pluchery,[d] Dominique Vuillaume,[b] Stéphane Lenfant,[b] Florence Volatron,[a] Anna Proust*[a]*

[a]Sorbonne Université, CNRS, Institut Parisien de Chimie Moléculaire, IPCM, 4 Place Jussieu, F-75005 Paris, France

[b] Institute for Electronics Microelectronics and Nanotechnology (IEMN), CNRS, Av. Poincaré, Villeneuve d'Ascq, France

[c]Sorbonne Université, CNRS, Laboratoire de réactivité de surface, LRS, 4 Place Jussieu, F-75005 Paris, France

[†] present address: FEMTO-ST Institute, UMR CNRS 6174, Université Bourgogne Franche-Comté, 15B avenue des Montboucons, 25030 Besançon Cedex, France

[d]Sorbonne Université, CNRS, Institut des Nanosciences de Paris, INSP, 4 Place Jussieu, F-75005 Paris, France

ABSTRACT

Immobilization of polyoxometalates (POMs) onto oxides is relevant to many applications in the fields of catalysis, energy conversion/storage or molecular electronics. Optimization and understanding the molecule/oxide interface is crucial to rationally improve the performance of the final molecular materials. We herein describe the synthesis and covalent grafting of POM hybrids with remote carboxylic acid functions onto flat Si/SiO2 substrates. Special attention has been paid to the characterization of the molecular layer and to the description of the POM anchoring mode at the oxide interface through the use of various characterization techniques, including ellipsometry, AFM, XPS and FTIR. Finally, electron transport properties were probed in a vertical junction configuration and energy level diagrams have been drawn and discussed in relation with the POM molecular electronic features inferred from cyclic-voltammetry, UV-visible absorption spectra and theoretical calculations. The electronic properties of these POM-based molecular junctions are driven by the POM LUMO (d-orbitals) whatever the nature of the tether or the anchoring group.

## INTRODUCTION

Polyoxometalates (POMs) are nanometric molecular oxides made of early transition metals in their highest oxidation state. Their ability to be reduced with several electrons with minor structural changes makes them good electron reservoirs that could play important roles in various applications in molecular memories, energy storage and energy conversion, catalysis…[1–9] Some of these applications involve the use of oxide electrodes. Indeed, in capacitive molecular memories, the molecular floating gate is usually deposited onto the silicon channel protected by silicon dioxide.[1,10,11] Thus, the molecules are not perturbed by the channel and keep their charge state and physical properties. Furthermore, in the field of solar energy conversion, *e.g.* photovoltaic and artificial photosynthesis (proton or $CO_2$ reduction, water oxidation...), nanostructured transparent metal oxide semiconductors like ITO, $TiO_2$, NiO or ZnO, and more recently the promising $Fe_2O_3$ or $BiVO_4$ semiconductors, often modified by an organic dye, are usually used to build photo-electrodes.[12–16]

Exploiting the redox activity of POMs in such applications thus requires mastering their deposition on oxide surfaces. Furthermore, the control of their organization on the surface is an essential and still challenging issue. Indeed, Busche et al., who recently reported the study of POM-based flash memories showed that the density of deposited POMs is crucial to control the write/erase voltage and switching time.[1] In the energy field, it is well known that the activity of dye-sensitized photo-electrodes will be maximized if the contact between the active molecules and the electrode is optimized, which implies: (i) limiting the aggregation of the molecules, (ii) preventing the leaching of the molecules during operating conditions, (iii) ensuring a strong interaction between the molecules and the electrode to densify the molecules at the interface and speed up electron transfers.[3,17–20] An efficient answer to all these points is to covalently attach the redox active molecules on the electrode.

Herein we demonstrate the robust and controlled grafting of POM hybrids onto $Si/SiO_2$ substrates. We chose this oxide for its direct applicability in CMOS (complementary metal-oxide

semi-conductor)-compatible devices, but also because very flat Si/$SiO_2$ substrates are available, which permits the use of numerous characterization techniques to probe the surface and interface. Whereas various strategies have been proposed to covalently anchor POMs on oxide nanoparticles surface[21–23] or mesoporous silica,[24] very few examples describe the grafting of POMs on flat oxide substrates.[13,25] In these last examples, few attention has been devoted to the control of the density and the characterization of the interface. In the present work, to ensure a strong bond between the POMs and the oxide, the POM hybrids with a pendant carboxylic acid group $TBA_{4.4}[PW_{11}O_{39}\{Sn(C_6H_4)C{\equiv}C(C_6H_4)COOH_{0.6}\}]$ ($K^{W}_{Sn}$[COOH]) and $TBA_{3.4}[PW_{11}O_{39}\{O(SiC_2H_4COOH_{0.8})_2\}]$ ($K^{W}_{Si}$[COOH]) have been prepared (TBA = $N(C_4H_9)_4^+$, tetrabutylammonium). Molecules containing carboxylic acid moieties were scarcely used to functionalize flat silicon oxide substrates[26,27] but the use of this anchoring function is recognized as a common way to modify non-planar silica surfaces and other oxides.[28–30]

Once synthesized, the POM hybrids were deposited on a silicon substrate surmounted by its native $SiO_2$ layer (hereafter named Si/$SiO_2$ substrate). Various surface characterization techniques were used to study the features of the monolayer and thorough attention was devoted to characterize the anchoring mode at the interface. To go further, the charge transport through the modified substrates was studied at the solid state.

## EXPERIMENTAL SECTION

All chemicals and solvents were purchased form Aldrich or Acros and used as received, except for triethylamine and acetonitrile that were distilled from $CaH_2$. The lacunar POM $K_7[PW_{11}O_{39}]$ and the hybrid platform $TBA_4[PW_{11}O_{39}\{SnC_6H_4I\}]$ ($K^{W}_{Sn}$[I], TBA = $N(C_4H_9)_4^+$, tetrabutylammonium) were synthesized following previously reported procedures.[31,32] The silicon wafers (highly phosphorus-doped n-Si(100), resistivity <$5.10^{-3}$ Ω.cm) for AFM, XPS and electrical characterizations were purchased from Siltronix and the silicon wafers for FTIR spectroscopy (Float-zone, low phosphorus-doped n-Si(100), resistivity 20-30 Ω.cm) were purchased from Neyco.

NMR spectra were recorded on a Bruker AvanceIII Nanobay 400 MHz spectrometer equipped with a BBFO probehead. $^{1}H$ chemical shifts are quoted as parts per million (ppm) relative to tetramethylsilane using the solvent signals as secondary standard (s: singlet, d: doublet, t: triplet, sex: sextet, m: multiplet) and coupling constants (J) are quoted in Hertz (Hz). $^{31}P$ chemical shifts are quoted relative to 85% $H_3PO_4$. IR spectrum of the powder was recorded from a KBr pellet on a Jasco FT/IR 4100 spectrometer. High-resolution ESI mass spectra were recorded using an LTQ Orbitrap hybrid mass spectrometer (Thermofisher Scientific, Bremen, Germany) equipped with an external ESI source operated in the negative ion mode. Spray conditions included a spray voltage of 3.5 kV, a capillary temperature maintained at 270 °C, a capillary voltage of −40 V, and a tube lens offset of −100 V. Sample solutions in acetonitrile (10 pmol.μL$^{-1}$) were infused into the ESI source by using a syringe pump at a flow rate of 180 μL.h$^{-1}$. Mass spectra were acquired in the Orbitrap analyzer with a theoretical mass resolving power ($R_p$) of 100 000 at m/z 400, after ion accumulation to a target value of $10^5$ and a m/z range detection from m/z 300 to 2000. All data were acquired using external calibration with a mixture of caffeine, MRFA peptide and Ultramark 1600 dissolved in Milli-Q water/ HPLC grade acetonitrile (50/50, v/v). Elemental analyses were performed at the Institut de Chimie des Substances Naturelles, Gif sur Yvette, France. Electrochemical studies were performed on an Autolab PGSTAT 100 workstation (Metrohm) using a standard three-electrode set-up. Glassy carbon electrode, platinum wire and saturated calomel electrode (SCE) were used as the working, auxiliary and reference electrode, respectively. The cyclic voltammograms were recorded in 1 mM solutions of the POMs in acetonitrile with tetrabutylammonium hexafluorophosphate $TBAPF_6$ as electrolyte (0.1 M). UV-visible spectra were recorded on a Cary 5000 spectrophotometer.

**Synthesis of $TBA_{4.4}[PW_{11}O_{39}\{Sn(C_6H_4)C{\equiv}C(C_6H_4)COOH_{0.6}\}]$ ($K^W_{Sn}[COOH]$)**

$K^{w}_{Sn}$[I] (200 mg, 0.050 mmol), 4-ethynylbenzoic acid (14.20 mg, 0.097 mmol), bis(triphenylphosphine) palladium (II) dichloride (2.83 mg, 0.004 mmol) and CuI (1.56 mg, 0.008 mmol) were solubilised in 2 mL of anhydrous DMF preliminary degassed with argon. Then triethylamine (50 µL, 0,378 mmol) was added. The mixture was degassed for 5 min more before being placed in a microwave oven for 1h at 80°C/80W. After the reaction, solid impurities were observed and removed by centrifugation. Diethyl ether was added to precipitate a powder that was solubilized again in a minimum of acetonitrile. The acetonitrile solution was stirred with $TBA^{+}$ enriched Amberlite® ion exchange resin for 1hr. The addition of few drops of ethyl acetate leads to the apparition of a preliminary precipitate that was removed by centrifugation since it contains catalysts residues. The supernatant was then precipitated by further addition of ethyl acetate, $K^{w}_{Sn}$[COOH] was recovered by centrifugation and dried with a large excess of diethyl ether to give a yellow pale powder (70-80%).

$^{1}$H NMR (400 MHz, $CD_3CN$): δ(ppm) 8.01 (d, $^{3}J_{H-H}$=8.02Hz, 2H, Ar-*H*), 7.75 (d+dd, $^{3}J_{H-H}$=7.74Hz, $^{3}J_{Sn-H}$= 93.4 Hz, 2H, Ar-*H*), 7.66 (d+dd, $^{3}J_{H-H}$=7.74Hz, $^{4}J_{Sn-H}$= 31 Hz, 2H, Ar-*H*), 7.60 (d, $^{3}J_{H-H}$=8.02Hz, 2H, Ar-*H*), 3.12 (m, 35H, N-C$H_2$-$CH_2$-$CH_2$-$CH_3$), 1.63 (m, 35H, N-$CH_2$-C$H_2$-$CH_2$-$CH_3$), 1.39 (sex, $^{3}J_{H-H}$=7.74Hz, 35H, N-$CH_2$-$CH_2$-C$H_2$-$CH_3$), 0.98 (t, $^{3}J_{H-H}$=7.74Hz, 53H, N-$CH_2$-$CH_2$-$CH_2$-C$H_3$); $^{31}$P NMR (121 MHz, $CD_3CN$): δ(ppm) – 10.90 (s+d, $^{2}J_{Sn-P}$=12.11Hz); IR (KBr pellet, $cm^{-1}$): ν=2961 (m), 2934 (m), 2873 (m), 1709(w), 1604(w), 1548 (w), 1483 (m), 1379 (w), 1070 (s), 962 (s), 886 (s), 798 (s); HRMS ($ESI^{-}$): m/z: calcd for $PW_{11}O_{41}SnC_{15}H_9$ : 754.30 $[M]^{4-}$ ; found 754.30 (100) ; calcd for $PW_{11}O_{41}SnC_{31}H_{45}N$: 1086.50 [M + TBA]$^{3-}$ ; found 1086.50 (50) ; calcd for $PW_{11}O_{41}SnC_{15}H_8$ : 603.24 $[M]^{5-}$; found 603.24 (25) ; calcd for $PW_{11}O_{41}SnC_{47}H_{81}N_2$ : 1750.89 [M + 2TBA]$^{2-}$; found 1750.89 (15); Anal. Calcd for $TBA_{4.4}[PW_{11}O_{39}\{Sn(C_{14}H_8)COOH_{0.6}\}]$ (%): C 25.12, H 4.12, N 1.51; found : C 24.80, H 3.92, N 1.54; UV-vis spectroscopy ($CH_3CN$): λ (nm) (logε, ε in $mol^{-1}.L.cm^{-1}$): 263 (4.73), 283 (4.66), 300 (4.69), 319 (4.59)

**Synthesis of $TBA_4[PW_{11}O_{39}\{Sn(C_6H_4)C{\equiv}C(C_6H_5)\}]$ ($K^{w}_{Sn}$[H])**

$K^{W}_{Sn}$[I] (400 mg, 0.1 mmol), ethynylbenzene (20 µL, 0.18 mmol), bis(triphenylphosphine) palladium (II) dichloride (7 mg, 0.01 mmol) and CuI (3 mg, 0.015 mmol) were solubilised in 4 mL of anhydrous DMF preliminary degassed with argon. Then triethylamine (100 µL, 0.756 mmol) was added. The mixture was degassed for 5 min more then left stirring under inert and ambient temperature during one night. TBABr (500 mg, 1.55 mmol) was added to the mixture then an excess of absolute ethanol to precipitate the product. A solid was recovered by centrifugation, washed with absolute ethanol and diethyl ether. To optimize the number of TBA counter-cations, the solid was solubilized in the minimum of acetonitrile, and stirred with $TBA^+$ enriched Amberlite® ion exchange resin for 1hr. After filtration, the filtrate was precipitated with ether and the solid recovered by a last centrifugation. $K^{W}_{Sn}$[H] was obtained with a 60% yield.

$^1$H NMR (400 MHz, $CD_3CN$): δ(ppm) 7.76 (d+dd, $^3J_{H-H}$=8.2Hz, $^3J_{Sn-H}$= 95 Hz, 2H, Ar-*H*), 7.65 (d+dd, $^3J_{H-H}$=8.2Hz, $^4J_{Sn-H}$= 32 Hz, 2H, Ar-*H*), 7.6 (m, 2H, Ar-*H*), 7.45 (m, 3H, Ar-*H*), 3.16 (m, 32H, N-C*H*$_2$-$CH_2$-$CH_2$-$CH_3$), 1.66 (m, 32H, N-$CH_2$-C*H*$_2$-$CH_2$-$CH_3$), 1.42 (sex, $^3J_{H-H}$=7.3Hz, 32H, N-$CH_2$-$CH_2$-C*H*$_2$-$CH_3$), 1.01 (t, $^3J_{H-H}$=7.3Hz, 48H, N-$CH_2$-$CH_2$-$CH_2$-C*H*$_3$); $^{31}$P NMR (121 MHz, $CD_3CN$): δ(ppm) – 10.9 (s+d, $^2J_{Sn-P}$=12.15Hz); IR (KBr pellet, $cm^{-1}$): ν=2962 (m), 2931 (m), 2869 (m), 1626(w), 1597(vw), 1483 (m), 1380 (w), 1069 (s), 961 (s), 885 (s), 798 (s); HRMS (ESI$^-$): m/z: calcd for $PW_{11}O_{39}SnC_{14}H_9$ : 743.30 $[M]^{4-}$ ; found 743.30 (100) ; calcd for $PW_{11}O_{39}SnC_{14}H_{10}$ : 991.41 $[M+H]^{3-}$ ; found 991.41 (70) ; calcd for $PW_{11}O_{39}SnC_{30}H_{46}N$ : 1608.25 $[M + H + TBA]^{2-}$ ; found 1608.26 (30) ; calcd for $PW_{11}O_{39}SnC_{30}H_{45}N$ : 1071.83 $[M + TBA]^{3-}$ ; found 1071.83 (20); calcd for $PW_{11}O_{39}SnC_{46}H_{81}N_2$: 1028.89 $[M + 2TBA]^{2-}$; found 1028.90 (15) Anal. Calcd for $PW_{11}O_{39}SnC_{78}H_{153}N_4$ (%): C 23.75, H 3.88, N 1.42; found: C 23.68, H 3.69, N 1.01; UV-vis spectroscopy ($CH_3CN$): λ (nm) (logε, ε in $mol^{-1}.L.cm^{-1}$): 271 (5.15), 278 (5.13), 285 (5.11), 304 (4.87).

**Synthesis of $TBA_{3.4}[PW_{11}O_{39}\{O\ (SiC_2H_4COOH_{0.8})_2\}]$ ($K^{W}_{Si}$[COOH])**

$K_7[PW_{11}O_{39}]$ (0.64 g, 0.2 mmol) was dissolved in a water/acetonitrile mixture (30 mL, 1:2). A 1 M HCl aqueous solution was added drop by drop until an apparent pH equals to 3. The solution was cooled in an ice bath and the $Si(OH)_3(CH_2)_2COONa$ (0.476 mL, 0.8 mmol) was inserted. The 1 M HCl solution was added drop by drop again to reach $pH_{app}$=2. After an overnight reaction, TBABr (0.26 g, 0.8 mmol) was added and the solution concentrated with a rotary evaporator to precipitate the product. The oily compound obtained was dissolved in the minimum of acetonitrile then precipitated again with an excess of ether. A sticky solid was recovered by centrifugation and washed thoroughly with ether to obtain a white powder (0.6 g, 82%).

$^1$H NMR (400 MHz, $CD_3CN$): δ(ppm) 3.14 (m, 27H), 2.53 (m, 4H), 1.65 (m, 27H), 1.41 (sex, $^3J$(H,H)=7.5 Hz, 27H), 1.05 (m, 4H), 1.01 (t, $^3J$(H,H)=7.5 Hz, 41H), 0.90 (m, 4H); $^{31}$P NMR (121 MHz, $CD_3CN$) δ (ppm) -12.28; IR (KBr pellet, $cm^{-1}$): ν =2963 (s), 2935 (m), 2874 (w), 1710 (s), 1623 (w), 1483 (s), 1471 (s), 1420 (w), 1381 (m), 1112 (vs), 1064 (vs), 1052 (s), 1036 (s), 964 (vs), 870 (vs), 824 (vs); HRMS (ESI-), m/z (%) : calcd for $W_{11}PSi_2O_{44}C_6H_{10}$ : 965.41 $[M]^{3-}$; found : 965.42 (100) ; calcd for $W_{11}PSi_2O_{44}C_{22}H_{46}N$ : 1569.26 $[M+TBA]^{2-}$; found: 1569.27 (50) ; calcd for $W_{11}PSi_2O_{44}C_{22}H_{45}N$: 1045.84 $[M + TBA]^{3-}$; found : 1045.84 (10) ; calcd for $W_{11}PSi_2O_{44}C_{38}H_{81}N_2$: 1569.27 $[M + 2TBA]^{2-}$; found 1569.26 (8); Elemental Analysis calcd for $TBA_{3.4}[PW_{11}O_{39}\{O(SiC_2H_4COOH_{0.8})_2\}]$ (%): C 19.50, H 3.55, N 1.28; found: C 19.51, H 3.50, N 1.24; UV-vis spectroscopy ($CH_3CN$): λ (nm) (logε, ε in $mol^{-1}.L.cm^{-1}$): 265 (5.59)

**Surface grafting**

The silicon substrate (1 $cm^2$ approximately) was first rinsed with a dichloromethane flow then dried under $N_2$. It was then treated by sonication in a basic piranha bath $NH_4OH/H_2O_2/H_2O$ (1/1/2 vol.) during 10 min. The treatment was repeated two more times, with deionized water rinse between each bath. The substrate was then rinsed by immersion in two successive deionized water baths under sonication during 5 min. After a thorough drying with $N_2$ flow, the substrate was immersed in an acetonitrile solution of the POM hybrid (1 $mmol.L^{-1}$) and heated at the reflux of the solvent during 22-24h. The modified substrate was rinsed following the following procedure: 5 min sonication in pure acetonitrile / rinsing with acetonitrile flow / 5 min sonication in a solution of 0.1 M of $TBAPF_6$ in acetonitrile / rinsing with acetonitrile flow / 5 min sonication in pure acetonitrile. The surface was finally rinsed with a last flow of acetonitrile before being dried with $N_2$. This rinsing sequence could be repeated once or twice depending on the ellipsometry results.

### Ellipsometry

Monowavelength ellipsometer SENTECH SE 400 equipped with a He-Ne laser at $\lambda$ = 632.8 nm was used to perform ellipsometry measurements. The incident angle was 70°. The values $n_s$ = 3.875 and $k_s$ = 0.018 were taken for the silicon wafer, $n_s$ = 1.5 and $k_s$ = 0 for the $SiO_2$ layer and $n_s$ = 1.48 and $k_s$ = 0 for the layer of POMs.[33,34] The thickness of the $SiO_2$ layer was estimated by doing a measurement by ellipsometry after the bare silicon substrate rinsing steps, just before immersing the substrate in the POM solution. The thickness of the POM layer was deduced by the subtraction of the $SiO_2$ layer thickness to the final thickness. At each step, 5 measurements were performed on the substrate in different zones, to check the homogeneity of the layer. A mean value for the thickness was calculated when the standard deviation was lower than 0.2 nm.

### Atomic Force Microscopy (AFM)

AFM images were recorded using a commercial AFM (NanoScope VIII MultiMode AFM, Bruker Nano Inc., Nano Surfaces Division, Santa Barbara, CA) equipped with a 150 × 150 × 5 µm scanner (J-scanner). The substrates were fixed on a stainless steel sample puck using a small piece of adhesive tape. Images were recorded in peak force tapping mode in air at room temperature (22−24 °C) using oxide-sharpened microfabricated $Si_3N_4$ cantilevers (Bruker Nano Inc., Nano Surfaces Division, Santa Barbara, CA). The spring constants of the cantilevers were measured using the thermal noise method, yielding values ranging from 0. 4 to 0.5 N/m. The curvature radius of silicon nitride tips was about 10 nm (manufacturer specifications). The raw data were processed using the imaging processing software NanoScope Analysis, mainly to correct the background slope between the tip and the surfaces.

## X-ray Photoelectron Spectroscopy (XPS)

XPS analyses were performed using an Omicron Argus X-ray photoelectron spectrometer. The monochromated $AlK_{\alpha}$ radiation source ($h\nu$ = 1486.6 eV) had a 300 W electron beam power. The emission of photoelectrons from the sample was analyzed at a takeoff angle of 90° under ultra-high vacuum conditions ($\leq 10^{-10}$ Torr). Spectra were carried out with a 100 eV pass energy for the survey scan and 20 eV pass energy for the C1s, O1s, N1s, Si 2p, P 2p, W 4f, Sn 3d, Pd 3d regions. Binding energies were calibrated against the Si2p binding energy at 99.4 eV and element peak intensities were corrected by Scofield factors. The spectra were fitted using Casa XPS v.2.3.15 software (Casa Software Ltd., U.K.) and applying a Gaussian/Lorentzian ratio G/L equal to 70/30.

## Fourier Transform Infra Red (FTIR) spectroscopy on silicon

FTIR spectra were recorded with a Bruker Tensor 27 spectrometer, equipped with a DTGS detector, at a resolution of 4 $cm^{-1}$. Detection was performed in transmission at an incident angle of 70° (Brewster angle for the air-silicon interface) in order to minimize the interferometric patterns in the spectra. Sample compartment is purged from water vapor with a commercial air-dryer (relative humidity around 7%). The silicon substrates are chosen to be float-zone grown and lightly doped in order to have a very low amount of inserted oxygen and to be transparent for the mid-IR wavelengths used in this study.

## I-V measurements

Current-voltage (I-V) curves were measured by contacting the POM monolayer by a Hg drop acting as the top electrode in a glove box filled under a nitrogen flow. The drop was gently brought into contact with the sample surface thanks to a camera. The voltage V was applied on the Hg drop and the highly doped silicon substrate is grounded through the ammeter to measure the current. I-V traces were acquired at different locations (around 10) on the POM monolayer surface, with ~ 5 traces per location, and they were not averaged. The surface contact area is estimated with the camera for each drop formed. Surfaces measured by this approach are comprised in the range 2 - 5 x $10^{-4}$ $cm^2$.

## RESULTS AND DISCUSSION

### Synthesis of carboxylic acid terminated POMs

The POM hybrid $TBA_{4.4}[PW_{11}O_{39}\{Sn(C_6H_4)C{\equiv}C(C_6H_4)COOH_{0.6}\}]$ (hereafter named $K^{W}_{Sn}$[COOH]) was synthesized (see experimental section) by post-functionalization of the hybrid platform $TBA_4[PW_{11}O_{39}\{SnC_6H_4I\}]$ ($K^{W}_{Sn}$[I]) via a Sonogashira coupling, following a procedure now well mastered in our group (scheme 1).[31]

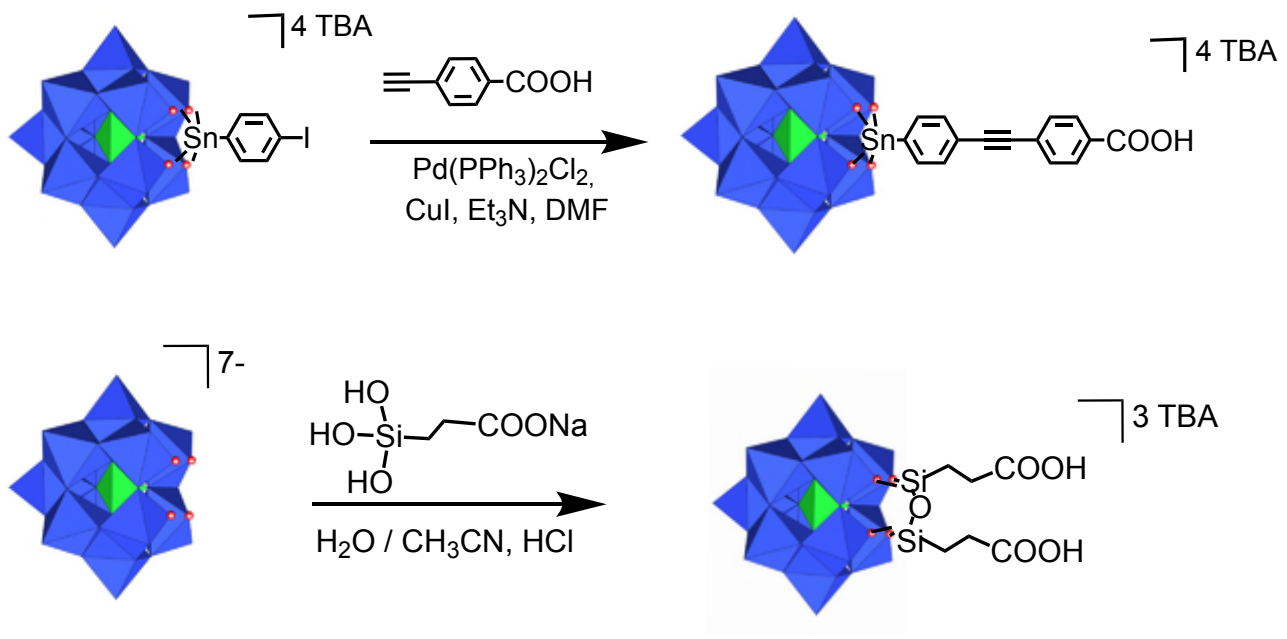

**Scheme 1.** Synthetic route to the carboxylic acid terminated hybrids $K^{W}_{Sn}$[COOH] (top) and $K^{W}_{Si}$[COOH] (bottom).

Fortunately, the coordinating carboxylic acid function of the reactant did not poison the catalyst and the POM hybrid was obtained as a yellow powder with a good yield (80%). $K^{W}_{Sn}$[COOH] was characterized by NMR and IR spectroscopies, mass spectrometry and elemental analyses (Figures S1-S4). On the $^1$H NMR spectrum, the two typical doublet pairs of the phenyl groups integrating for 2 protons appear at δ=7.74 and δ=7.66 ppm for the phenyl group linked to the inorganic core of the POM, and at δ=8.01 and δ=7.60 ppm for the phenyl group modified with the carboxylic acid function. The peaks corresponding to the tetrabutylammonium (TBA) cations are also present but the integration of the peaks does not correspond to 4TBA counter cations as expected with a pure carboxylic acid POM hybrid. More TBA cations are present around the POM hybrid, which means that a partial amount of POMs is in the carboxylate form.

This is confirmed by the IR study that shows, in addition to the band at 1709cm$^{-1}$ corresponding to the stretching vibration of the C=O bond of carboxylic acid function, a band at 1548 cm$^{-1}$ that can be attributed to the asymmetric vibration of a carboxylate group. The corresponding symmetric vibration band expected around 1420-1335 cm$^{-1}$ is probably hidden in the TBA bands at 1480 and 1380 cm$^{-1}$. The carboxylate form is confirmed by the observation of the non-majority peak at m/z = 603.24 on the mass spectrum, interpreted by the presence of a species charged five minus of formula $C_{15}H_8O_{41}SnPW_{11}$. Finally, elemental analysis complies with the following formula: $TBA_{4.4}[PW_{11}O_{39}\{Sn(C_6H_4)C{\equiv}C(C_6H_4)COOH_{0.6}\}]$ from which we can conclude that in the powder, 30% of the molecules are in the carboxylate form whereas 70% are in the carboxylic acid form.

The cyclic voltammogram (CV) of $K^{W}_{Sn}$[COOH] ($10^{-3}$ M) was recorded in a 0.1M $TBAPF_6$ / acetonitrile solution. The CV recorded at 100 mV.s$^{-1}$ shows three reversible reduction waves at $E_{1/2}$=-1.01 ($E_1$), -1.48 ($E_2$) and -1.64 V ($E_3$) vs saturated calomel electrode (SCE). The value of the two first reduction processes are in accordance with that reported for other tin derivatives in this family of polyoxotungstates (figure 1) and correspond to successive W(VI) to W(V) one-electron reduction processes.[35]

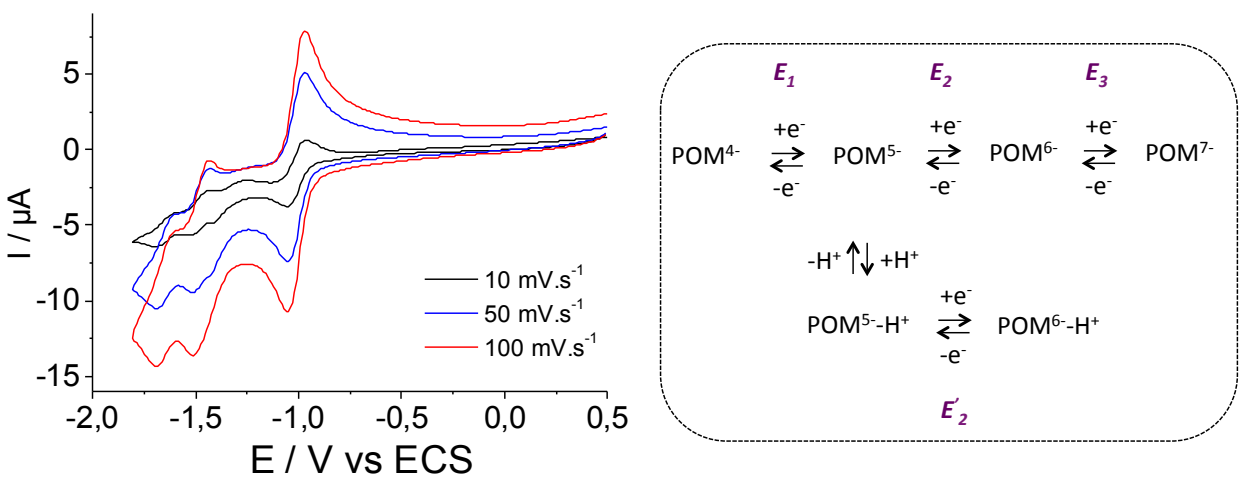

**Figure 1**. (left) cyclic voltammogram of $K^{W}_{Sn}$[COOH] in acetonitrile with $TBAPF_6$ 0.1 M at various scan rates ; 100 mV.s$^{-1}$ (red curve) ; 50 mV.s$^{-1}$ (blue curve) ; 10 mV.s$^{-1}$ (black curve) ; (right) scheme of the mechanism of successive electron and proton transfers. $POM^{4-}$= $[PW_{11}O_{39}\{Sn(C_6H_4)C{\equiv}C(C_6H_4)COOH\}]^{4-}$ (potentials are referenced to saturated calomel electrode, add 382 mV to convert to the $Fc^{+/0}$ scale.[36]

However, on the second reduction wave, a slight inflexion is observed. To better understand its origin we recorded CVs at various scan rates and observed that: at low scan rate (50 mV.s$^{-1}$ and below), the inflexion becomes a full wave at around -1.4 V vs SCE ($E'_2$, figure 1); at a scan rate higher than 100 $mV.s^{-1}$, the inflexion disappears (figure S5). We propose that after the first reduction process the formed "$POM^{5-}$" becomes more basic and undergoes intra or intermolecular protonation from a carboxylic acid group. The protonated "$POM^{5-}$-$H^{+}$" is thus more easily reduced and the wave $E'_2$ corresponds to the reduction potential of a portion of POMs that are protonated (figure 1).[37] The second reduction wave of this "$POM^{5-}$-$H^{+}$" species is probably hidden in the second or third reduction waves of the unprotonated "$POM^{4-}$". As the protonation reaction is slower than the electron transfer,[38] a portion of the "$POM^{5-}$" have time to be protonated then reduced at low scan rate whereas at a scan rate high enough, only the reduction of the unprotonated "$POM^{5-}$" occurs.

The POM hybrid $TBA_{3,4}[PW_{11}O_{39}\{O(SiC_2H_4COOH_{0,8})_2\}]$ (hereafter named $K^{W}_{Si}$[COOH]) was also synthesized (see experimental section), by adapting a published procedure.[39] This POM is quite different from the precedent since it displays aliphatic tethers, two anchoring groups and a total charge of -3 that should induce different electrochemical properties. It was prepared by a condensation of the carboxyethylsilanetriol on the lacunar POM $K_7[PW_{11}O_{39}]$ (figure 1) in acidic acetronitrile-water. NMR, infrared spectroscopies and mass spectrometry confirmed the obtaining of the expected species, in particular by the presence on the $^{1}H$ NMR spectrum of both multiplets at 1.06 and 2.53 ppm integrating for 4 protons corresponding to the protons on the aliphatic chains of the POM hybrid (figure S6-S9). The excess of TBA cations measured by NMR spectroscopy and two contributions in the mass spectrometry spectrum again shows that deprotonated species are present, which is confirmed by elemental analysis that complies with the following general formula: $TBA_{3,4}[PW_{11}O_{39}\{O(SiC_2H_4COOH_{0,8})_2\}]$. On the CV of the $K^{W}_{Si}$[COOH] recorded in a 1 mM solution in acetonitrile with $TBAPF_6$ (0.1M), the two first one electron reduction waves can be observed at $E_{1/2}$=-0.38 and -0.88 V versus SCE in accordance with other Keggin type silicon derivatives reported in the literature (figure S10).[40] As expected, the $K^{W}_{Si}$[COOH], less charged, is easier to reduce than the $K^{W}_{Sn}$[COOH].

### UV-visible spectra and DFT calculations

The UV-visible spectra of $K^{W}_{Si}$[COOH], $K^{W}_{Sn}$[COOH] and the reference compound (without any functional group) $TBA_4[PW_{11}O_{39}\{Sn(C_6H_4)C{\equiv}C(C_6H_5)\}]$ ($K^{W}_{Sn}$[H]), recorded in acetonitrile solution, are shown on Figure 2. All display the oxygens to metal(VI) charge transfer bands characteristic of POMs around 265 nm and below. In the case of $K^{W}_{Sn}$[COOH] and $K^{W}_{Sn}$[H] additional absorptions are observed at lower energies and are related to the conjugated organic chain. All together these electronic absorptions disclose the presence of a set of low-lying empty orbitals, which might be involved in the charge transport (see below).

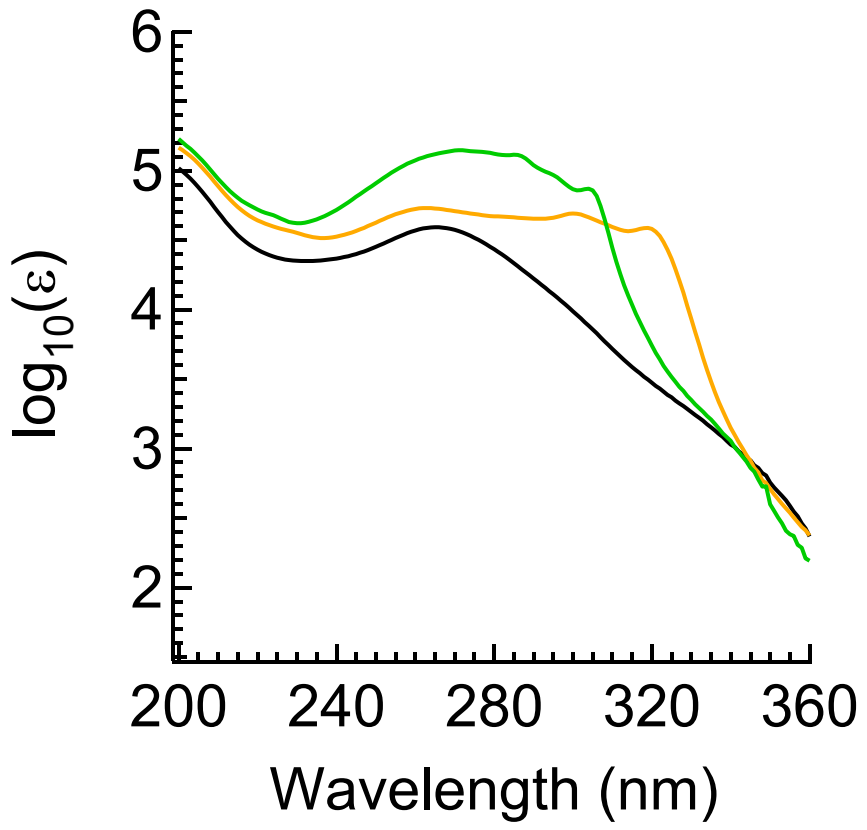

**Figure 2**. UV visible spectra of $K^{W}_{Si}$[COOH] (black), $K^{W}_{Sn}$[COOH] (orange) and $K^{W}_{Sn}$[H] (green), recorded in acetonitrile.

To gain more insights into their relative energies and the nature of the allowed electronic transitions, DFT and TD-DFT calculations have been undertaken, using the Turbomole package, version 6.4.[41] First of all, two model systems were selected: one corresponding to $K^{W}_{Si}$[COOH] and one corresponding to $K^{W}_{Sn}$[COOH]. In the case of $K^{W}_{Sn}$[COOH], we chose to remove the carboxylic acid on the organic moiety for performing the calculations since the protonation state is unclear and we took $K^{W}_{Sn}$[H] as a model compound. Geometrical optimization for both systems was conducted at the B3LYP-D3/def2-SV(P) level, with solvation taken into account by adding a continuum solvation model (namely COSMO, with a dielectric constant of 48). A summary of the frontier orbitals for the two calculated systems can be found in Figures 3 and 4 for $K^{w}_{si}$[COOH] and $K^{w}_{Sn}$[H] respectively. For $K^{w}_{si}$[COOH], its orbital diagram is characterized by the following: the HOMO is localized on the sigma system of the two carboxylic acid interacting one with each other (in blue boxes on Figure 3). Four relatively similar orbitals of this type can be found (HOMO, HOMO-1, HOMO-3, HOMO-5). Another block of occupied frontier orbitals can be characterized by oxygens' lone pairs located on the polyoxometalate part of the system (red boxes on Figure 3). The lowest unoccupied molecular orbitals (LUMOs) are characterized by tungsten d orbitals of the polyoxometalate (green boxes on Figure 3). Just above these orbitals can also be found antibonding π orbitals corresponding to the WO oxo bond (purple boxes on Figure 3).

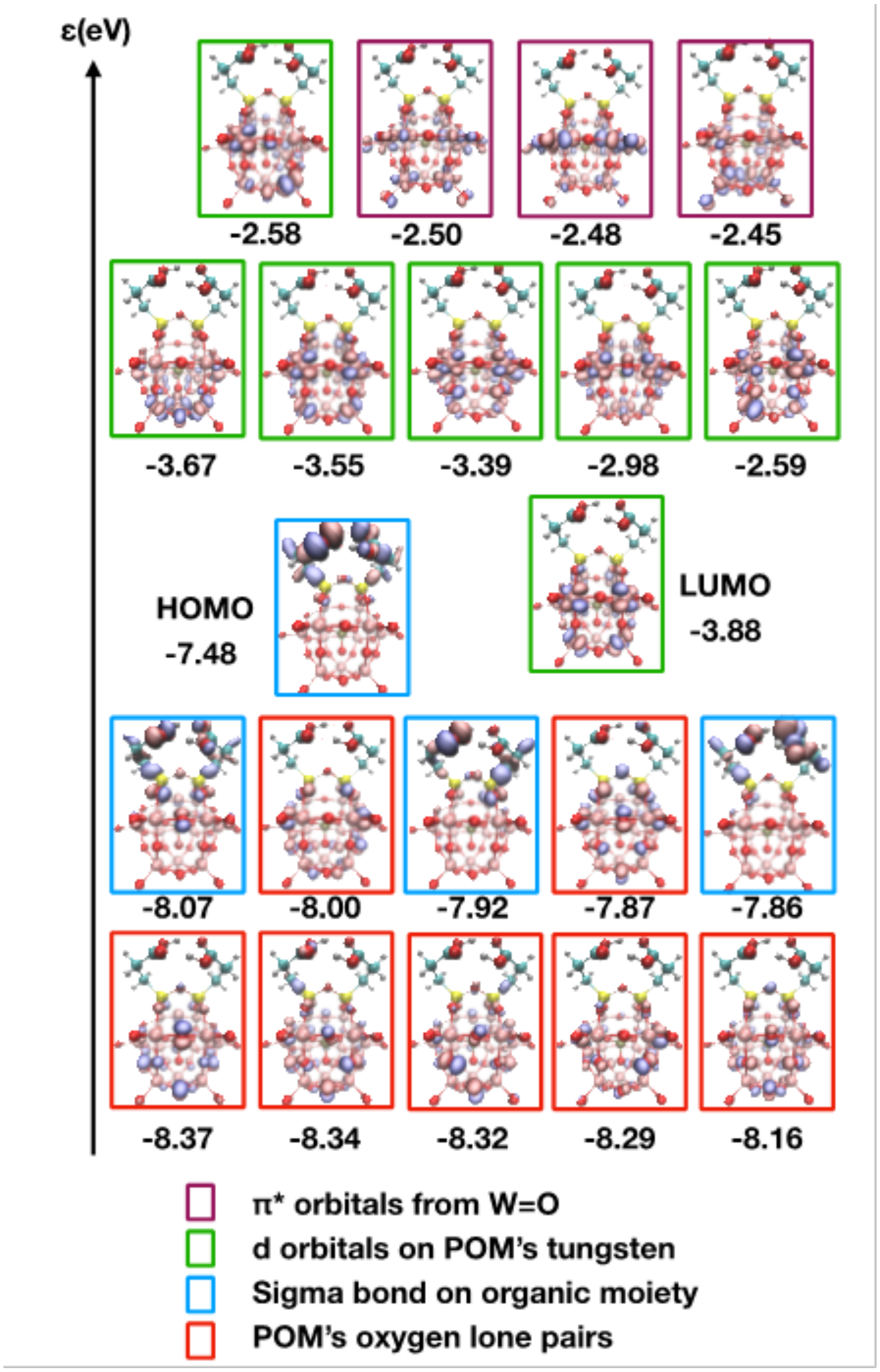

**Figure 3.** Frontier molecular orbitals for $K^{W}_{Sn}$[COOH], calculated at the B3LYP-D3/def2-SV(P). Color boxes indicates similar types of orbitals.

Regarding $K^{W}_{Sn}$[H], the set of frontiers orbitals is slightly different. HOMO to HOMO-4 can be easily characterized as π orbitals localized on the organic moiety of the hybrid (black boxes on Figure 4). Just below these five orbitals, one can find the usual oxygens' lone pairs located on the polyoxometalate part of the system (red boxes on Figure 4). The lowest unoccupied orbitals for $K^{w}_{sn}$[H] are similar to those of $K^{w}_{si}$[COOH]: first a block of six orbitals corresponding to combination of tungsten d orbitals (green boxes on Figure 4) followed by antibonding π orbitals corresponding to the oxo bond between tungsten and oxygen (purple boxes on Figure 4).

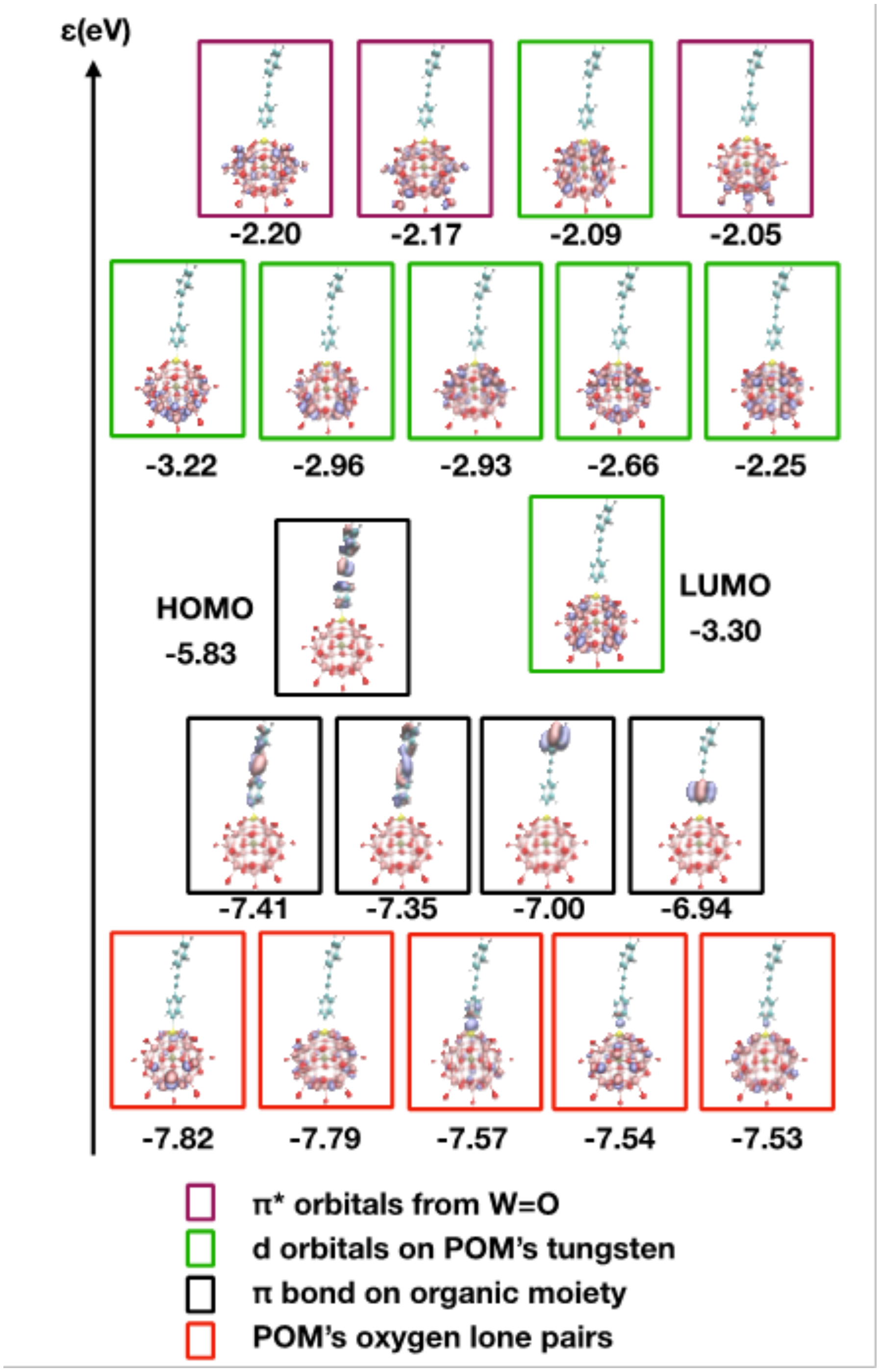

**Figure 4.** Frontier molecular orbitals for $K^{W}_{Sn}$[H], calculated at the B3LYP-D3/def2-SV(P). Color boxes indicates similar types of orbitals.

TD-DFT calculations were then performed to obtain electronic transition attribution. As seen experimentally, the UV spectrum of $K^{W}_{Si}$[COOH] present a major band around 265 nm. This band is properly reproduced by TD-DFT calculations (band A on Figure 5) and is attributed as expected to an electronic transition between oxygens' lone pair towards metal d orbitals. In the case of $K^{W}_{Sn}$[H], the same band can be found (band B on Figure 5) involving similar orbital contributions. But, on top of that, we can also attribute the band observed experimentally around 320 nm. From our calculations, it appears that this band can be attributed to a transition between the $\pi$ system localized on the organic moiety of $K^{W}_{Sn}$[H] and d orbitals of POM's tungstens (band C on Figure 5). Because it involves the $\pi$ system, the position of this band is expected to be sensitive to the functionalization of the aryl unit, as observed experimentally (see Figure 2). The intensity of the calculated band is stronger than expected, which can be attributed to an overestimation of the oscillator strength involving extended conjugated system as previously noted by Tozer.[42]

In conclusion, the computational study gives some insights about the assignment of the main electronic transitions and confirms that the lowest unoccupied molecular orbitals are combination of tungsten d orbitals without any participation of the organic tether.

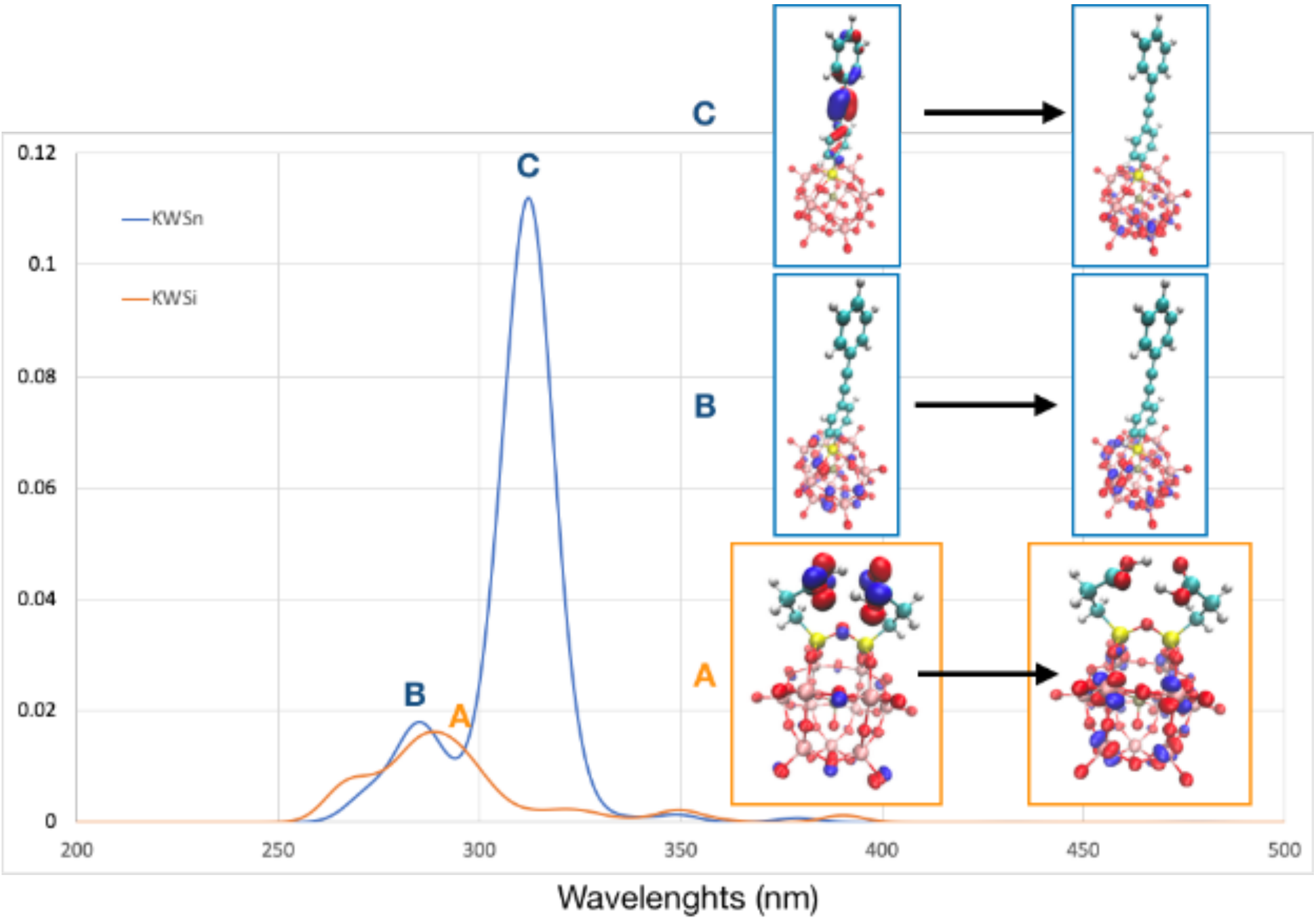

**Figure 5**. Theoretical UV-visible spectra for $K^{W}_{Sn}$[H] and $K^{W}_{Si}$[COOH], calculated at the B3LYP-D3/def2-SV(P) level. Inserts: main orbital contribution for selected electronic transitions.

## Grafting on silicon oxide

As mentioned in the introduction, whereas a huge literature exists about the deposition of POM films on oxide substrates by layer-by-layer,[16,43,44] spin coating,[15,45] or dip coating on positively charged organic monolayers,[19,46] involving Van der Waals or electrostatic interactions between the POMs and the substrate, only two examples reported the covalent bonding of POMs on flat oxide substrates. M. Tountas et al. described the insertion of *tert*-butyl silanol functionalized trivacant Keggin-type POMs $(TBA)_3[PW_9O_{34}(t\text{BuSiOH})_3]$ in a polymer solar cell at the metal oxide ($TiO_2$ or ZnO) / organic interface.[25] Interestingly, the authors proposed that strong POM-Si-O-metal bonds were formed at the oxide surface via the silanol groups inducing the formation of more homogeneous films and the passivation of the metal oxide surface stabilizing the system. This new type of interaction at the interface contributed to the improvement of the device. E. A. Gibson and J. Fielden and coll. reported the study of pristine and organo-imido Lindqvist-type POM hybrids (with carboxylic acid and pyridine pendant group) as co-adsorbents in dye sensitized p-type NiO solar cells.[13] They showed that the presence of the POMs at best slightly improves the overall cell efficiency despite a significant increase of $V_{oc}$. This was attributed to the combination of opposite effects such as retardation of both recombination to NiO and electron transfer to the electrolyte. Furthermore, no clear effect of the nature of the POM was observed. Specially, we could expect the POM hybrids with a pendant carboxylic acid function to induce a different behavior because of a strong anchoring to the NiO surface, albeit no experimental evidence was provided. One possible explanation is that the covalent grafting is not guaranteed in this study: the POMs layer is indeed added in a second step after prior coverage of the substrate by the dye sensitizer, also with a carboxylic acid anchor, which could hinder the subsequent interaction between the POMs and the oxide substrate.

Here, we chose to deposit the $K^{W}_{Sn}$[COOH] and $K^{W}_{Si}$[COOH] POM hybrids on $Si/SiO_2$, a very flat substrate, to have access to crucial surface characterization techniques, such as ellipsometry and AFM, to gain insights into the POM layer features. According to the literature, to ensure a strong bonding between the carboxylic acid and the oxide surface, an annealing step is required.[28] Thus in the present case, the deposition was done by heating at the solvent reflux an acetonitrile solution of the POMs containing the freshly cleaned $Si/SiO_2$ substrate for 24h. The substrate was then rinsed thoroughly, by sonication baths in pure acetonitrile but also in a solution of $TBAPF_6$ in acetonitrile to eliminate any POM electrostatically deposited.

The substrate with the $K^{W}_{Sn}$[COOH] layer was first characterized by ellipsometry. Theoretically, thicknesses between 2.7 and 3.6 nm are expected (figure 6a). Indeed, the length of the rigid organic arm is around 1.6 nm, the POMs size is 1 nm and the size of the flexible TBA counter cation can vary between 0.5 and 1 nm, as a function of its folding. If we consider molecules standing up-right in the normal direction to the surface with extended TBA, we expect a maximal thickness of 3.6 nm. If tilted molecules with the classical tilt angle of 30° are considered, the minimal expected thickness is 2.7 nm. Based on our statistics (figure S11), we defined 2.3 nm as the lower acceptable limit. Over 17 samples, a majority (60%) gave satisfying thicknesses (between 2.3 and 3.9 nm, Figure S11 in the supporting information)) and were regarded as monolayers. The samples obtained with a thickness below 2.3 nm and above 4 nm were regarded as sub-monolayers and multilayers respectively and were sidelined.

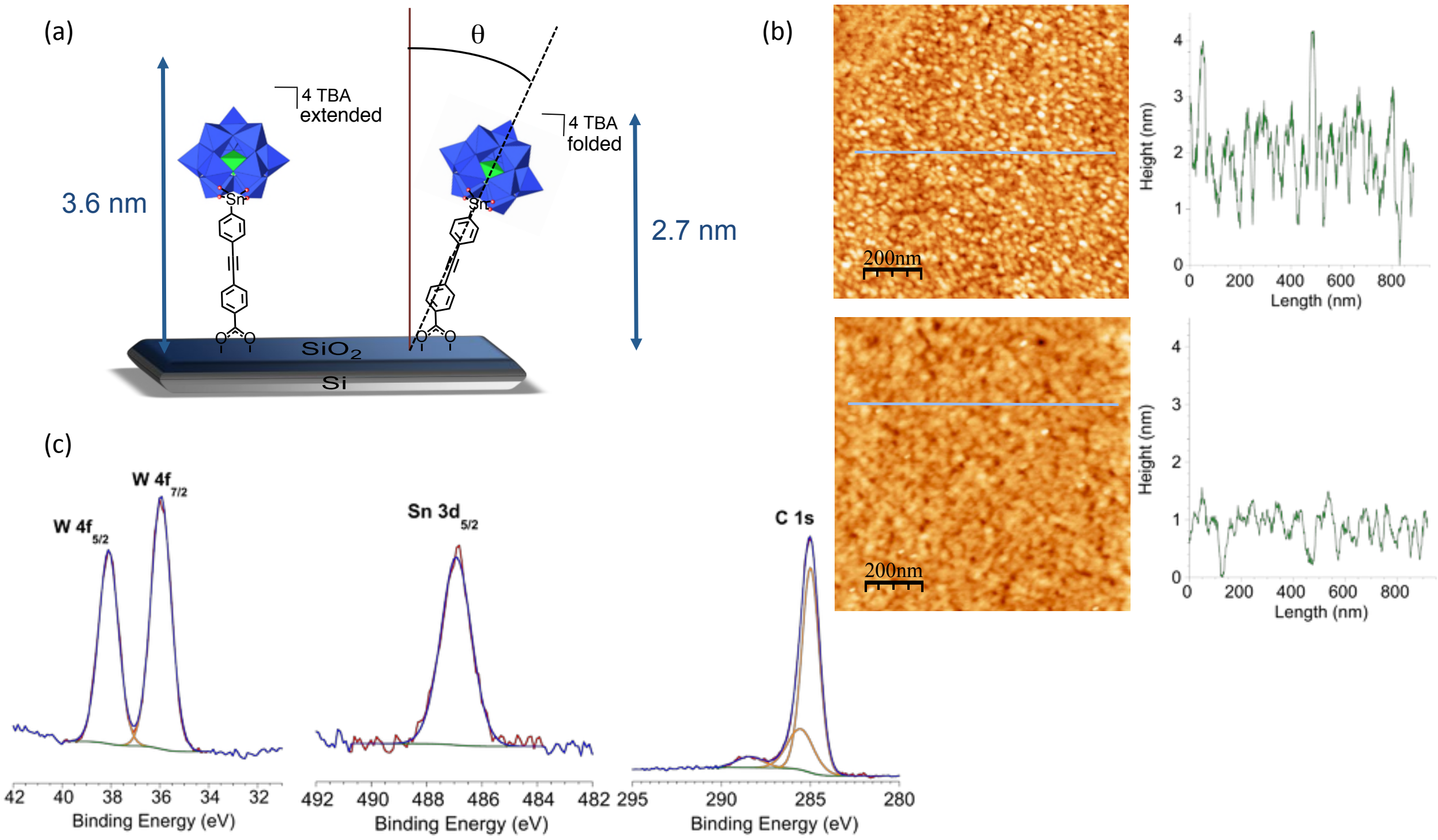

**Figure 6.** (a) Scheme of the extreme orientations of the $K^{W}_{Sn}$[COOH] layer on Si/$SiO_2$ (b) 1*1 µm AFM images and z-profile of the $K^{W}_{Sn}$[COOH] layer (top) compared to the freshly piranha treated Si/$SiO_2$ substrate (bottom) (c) W 4f, Sn 3d and C 1s high resolution XPS spectra of the $K^{W}_{Sn}$[COOH] layer on Si/$SiO_2$.

AFM images show that the layer is rougher (RMS roughness= 0.59 nm) than the bare Si/$SiO_2$ substrate (RMS roughness= 0.37 nm) but still very homogeneous (figure 6b and S13), *i.e.* we do not observe large defects like holes or clusters. XPS analysis permitted to detect all the POM elements on the substrate : the broad P 2p photopeak, the typical W $4f_{7/2}$ and $4f_{5/2}$ doublet at 36.0 and 38.1 eV respectively, and the Sn $3d_{5/2}$ photopeak at 486.8 eV, all these values corresponding to binding energies of the elements with oxygen neighbors (figure 6c and S14).[47–50] Carbon and nitrogen were also detected on the surface. The C 1s photopeak can be deconvoluted with three contributions: the one at 285.0 eV is attributed to aliphatic and aromatic C-C bonds as well as Sn-C bonds, the one at 285.5 eV corresponds to the carbons linked to the positively charged ammonium atom in the TBA counter-cations, and the last peak at 288.4 eV, widely shifted to higher binding energies, refers to the carbon linked to oxygen atoms in the terminal carboxylic group (figure 6c). Only one peak appears on the high-resolution spectrum of the N1s, corresponding to the nitrogen of the ammonium counter-cations (figure S14). Surprisingly, a significant quantity of palladium at the +2 charge state is observed on the survey and the Pd high resolution spectra (figure S15). The only rational explanation is to attribute this impurity to the palladium-based catalyst used in the Sonogashira coupling. However, XPS measurements performed on the $K^{W}_{Sn}$[COOH] powder show the absence of the Pd element (on the survey spectrum as well as on the high resolution spectrum performed at the Pd level, figure S15). This means that palladium is present as traces only in the powder but probably accumulates on the substrate. Moreover, on all the other substrates we have studied until now, made of similar POMs layers based on tin derivative POM hybrids that had been obtained by a palladium catalyzed Sonogashira coupling, but grafted on different substrates (carbon, gold, hydrogenated silicon),[33,47,51] traces of palladium had never been detected. The $SiO_2$ layer has probably a strong affinity with palladium and during the POMs deposition the palladium traces accumulate on the substrate. The $Pd_{5/2}$ photopeak at 338.0 eV confirms this hypothesis as it is in accordance with Pd species in an oxygenated environment.[52] Note that in a blank sample formed from the non-acidic POM $K^{W}_{Sn}$[H]), $Pd^{2+}$ is also detected (figure S16), supporting its strong interaction with $SiO_2$. Furthermore, the proportion of Pd compared to the W element is lower (20%, compared to 75% for the $K^{W}_{Sn}$[COOH] layer, see table 1), which means that the carboxylic acid group is also a drainage source of the palladium traces.

The substrate functionalized with $K^{W}_{Si}$[COOH] was also characterized. The highest theoretical thickness was evaluated to be at 2.6 nm with a 0.6 nm height for the organic tether, 1 nm for the inorganic core of the POM and 1 nm for the extended TBA counter-cations. The lowest limit for the thickness is more delicate to determine theoretically as two more parameters have to be taken into account: the flexibility of the tether, and the fact that the POM can be attached by one or two anchoring groups (figure 7a).

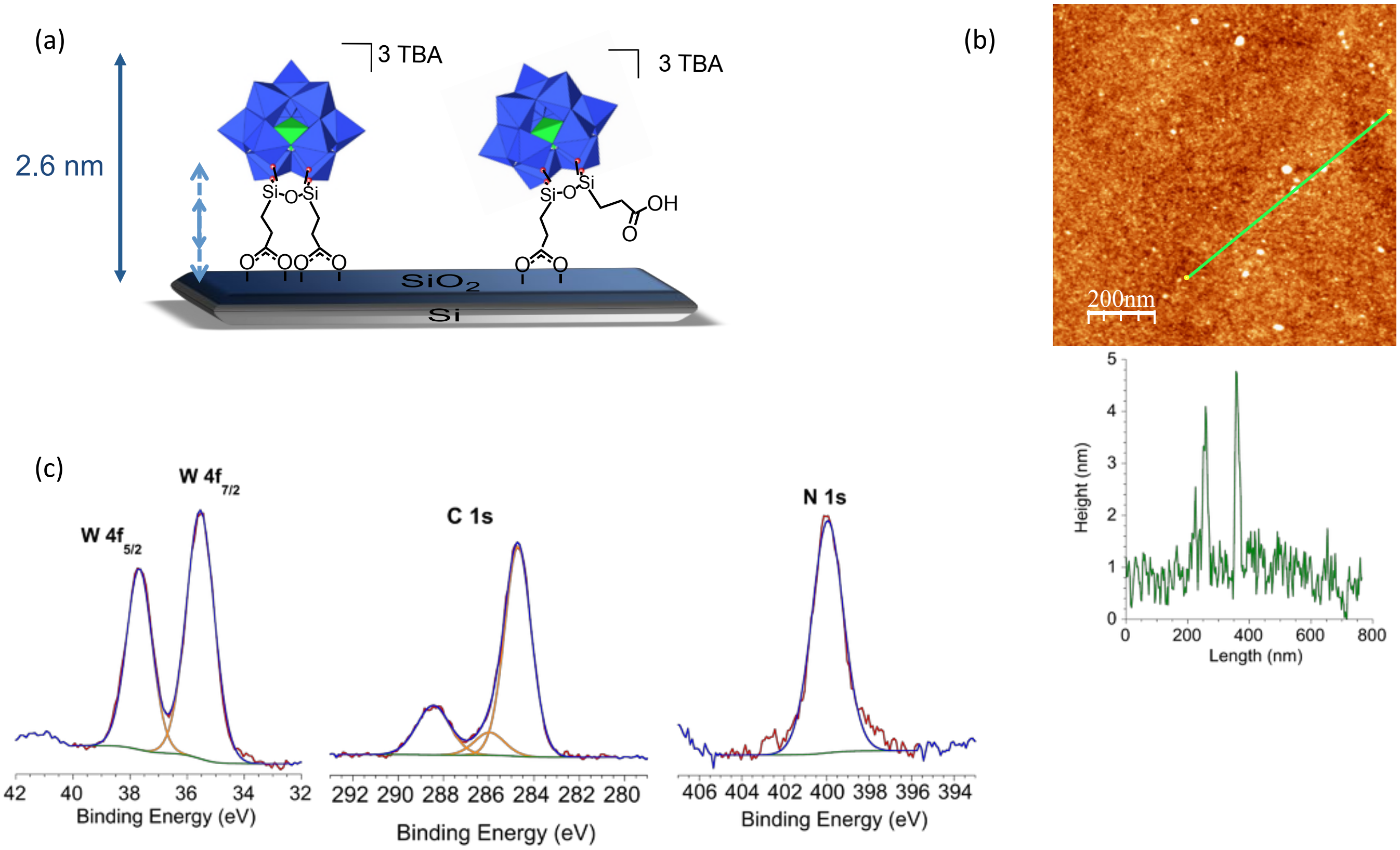

**Figure 7**. (a) Scheme of the possible anchoring modes of the $K^{W}_{Si}$[COOH] POMs on Si/$SiO_2$ (b) 1*1 µm AFM image and z-profile of the $K^{W}_{Si}$[COOH] layer (c) W 4f, C 1s and N 1s high resolution XPS spectra for the $K^{W}_{Si}$[COOH] layer on Si/$SiO_2$.

On 7 samples, 6 gave satisfactory results by ellipsometry (between 1.5 and 2.5 nm, figure S12). Nevertheless, the $K^{W}_{Si}$[COOH] layer is less homogeneous than the $K^{W}_{Sn}$[COOH] layer, as showed by the AFM images. The layer has an overall roughness of 0.4 nm, but with some aggregates with a size around 5-8 nm are observed on the 1*1 mm image (figure 7b), and seem regularly dispersed on the surface, according to the 5*5 mm image (figure S13). The flexibility of the anchoring mode, due to the tether nature and the two carboxylic acid groups may induce more disorder in the layer and explains the totally different behavior of $K^{W}_{Si}$[COOH] compared to $K^{W}_{Sn}$[COOH]. XPS measurements confirmed the presence of the constitutive elements of the POM (Figure 7c and S17). The high-resolution spectrum in the W4f bonding energy range shows two peaks at 35.5 and 37.3 eV corresponding to $W4f_{7/2}$ and $W4f_{5/2}$ respectively. The N 1s high-resolution spectrum shows as expected a unique peak at 400 eV proving the presence of the TBA counter-cations. As previously, the C 1s photopeak can be deconvoluted with 3 contributions at 284.7, 286 and 288.5 eV corresponding respectively to C-C / C-Si bonds, C-N+ bonds and C-O / C=O bonds. It is worth noting that the contribution of the carbon element linked to oxygen atoms is more important in the present case than in the case of the $K^{W}_{Sn}$[COOH] layer, in accordance with the two anchoring groups present in $K^{W}_{Si}$[COOH]. Moreover, as expected, no palladium was detected in this sample, confirming our previous explanations about the origin of the Pd (Figure S17).

**Insights into the anchoring mode** The three surface characterization techniques (AFM, XPS and ellipsometry) have thus proved that we were able to build monolayers of POMs. To go further, we tried to probe more specifically the $SiO_2$/POMs interface to elucidate the grafting mode of the POMs on the $SiO_2$ layer and particularly the exact nature of the bond between the carboxylic acid group and the $SiO_2$ layer. Indeed, various binding modes can occur between a carboxylic acid group and an oxide surface:[28] the bidentate carboxylate mode (chelating one silicon of the surface or bridging two adjacent silicon atoms of the surface), the monodentate carboxylate mode which results in the formation of an ester function, and the hydrogen bond mode, in which the carboxylic acid group forms H-bonds with the hydroxyl groups of the surface (see figure 8).

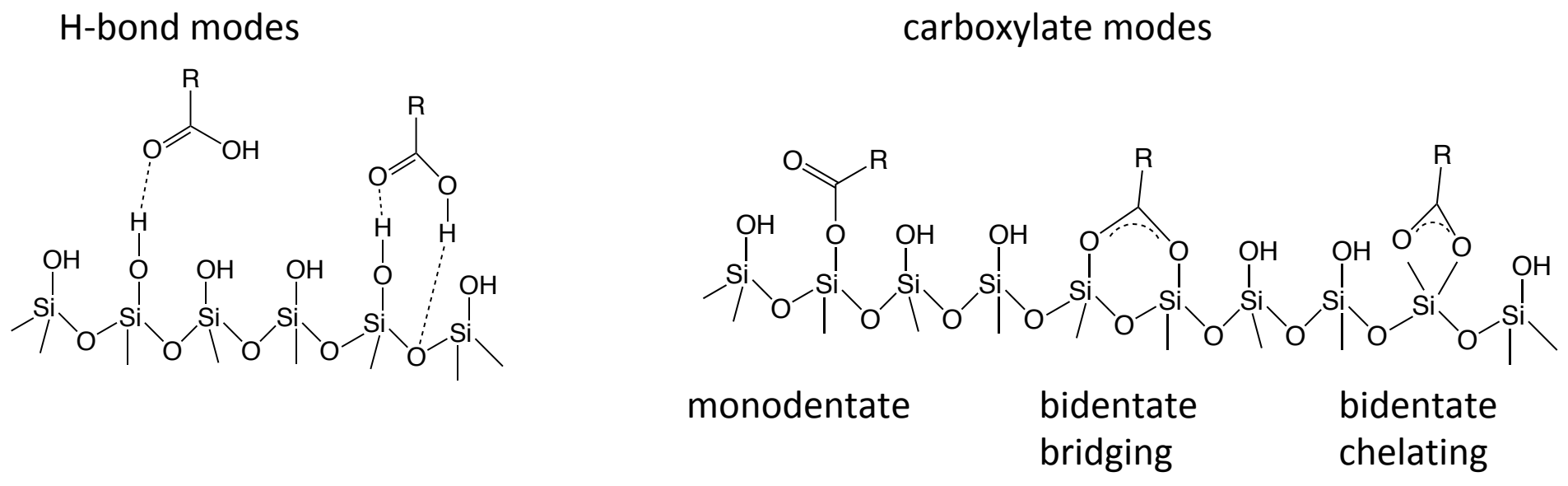

**Figure 8.** Possible binding modes of carboxylic acid or carboxylate groups to silicon oxide surfaces.[28]

Note that the rinsing step supports a covalent grafting: indeed, despite a thorough rinsing, which consist in treating the POMs modified substrate in a $TBAPF_6$ solution by an ultrasonic bath, the POMs layer keeps attached, revealing a strong anchoring with the substrate. To confirm this observation, a blank sample was prepared by immersing a $Si/SiO_2$ substrate in an acetonitrile solution of $K^{W}_{Sn}$[H], following exactly the same procedure as for the $K^{W}_{Sn}$[COOH] layer. The thickness for the $K^{W}_{Sn}$[H] layer measured by ellipsometry after thorough rinsing is only 0.5 nm, showing that only a few POMs are deposited on the surface. XPS measurements confirm this observation, as the atomic percentage ratio W/Si is around 25 times less important for the $K^{W}_{Sn}$[H] layer than for the $K^{W}_{Sn}$[COOH] layer (figure S16, Table 1).

| %at | $K^{W}_{Sn}$[COOH] layer | $K^{W}_{Sn}$[H] layer |
|---|---|---|
| W | 0.99 | 0.15 |
| Si | 13.27 | 49.13 |
| Pd | 0.74 | 0.03 |
| W/Si | $7.4.10^{-2}$ | $3.10^{-3}$ |
| Pd/W | 0.75 | 0.2 |

**Table 1**. Atomic percentages of W and Si elements in $K^{W}_{Sn}$[COOH] and $K^{W}_{Sn}$[$C_6H_5$] layers on $Si/SiO_2$

The $K^{W}_{Sn}$[H] layer is thus constituted of a few POMs, haphazardly physisorbed and lying down on the $Si/SiO_2$. The carboxylic acid function is thus crucial to ensure a vertical and robust anchoring of the POMs on the substrate, which is an indirect proof that the POMs are attached via their terminal group.

To go further and try to determine the nature of the link between the POMs and the surface, FTIR measurements were performed on both carboxylic acid POMs layers, deposited on a special low-doped float-zone silicon substrate, transparent to IR radiation. To validate the technique, a drop-casting of the powders was performed on a freshly cleaned $Si/SiO_2$ substrate (Figure S18). In both cases, the FTIR spectrum is similar to the one registered on a KBr pellet, with the typical bands of the inorganic core of the POM below 1150 $cm^{-1}$ and the vibration bands characteristic of the TBA counter-cations around 2900 $cm^{-1}$, and at 1382, 1467 and 1485 $cm^{-1}$. The band corresponding to water molecules of crystallization is also present in both samples, at 1675 $cm^{-1}$ for the $K^{W}_{Sn}$[COOH] drop-casted powder and at 1656 $cm^{-1}$ for the $K^{W}_{Si}$[COOH] drop-casted powder. More specifically on the $K^{W}_{Sn}$[COOH] spectrum, the C=O stretching band appears at 1704 $cm^{-1}$, and the aromatic C=C stretching band at 1606 $cm^{-1}$. The band corresponding to the asymmetric vibration of the carboxylate function can be guessed at 1548 $cm^{-1}$. On the $K^{W}_{Si}$[COOH] spectrum, the C=O stretching vibration band of the carboxylic acid group is far more intense at 1710 $cm^{-1}$. The band at 1602 $cm^{-1}$ can be attributed to the asymmetric stretching vibration of the carboxylate group, visible in this case thanks to the slight shift of the band corresponding to the crystallization water molecules toward 1656 $cm^{-1}$ compared to the spectrum of the powder registered in a KBr pellet.

The spectra of the POMs monolayers show different features, particularly in the 1300-1700 $cm^{-1}$ zone (Figure 9). On the spectrum of the $K^{W}_{Sn}$[COOH] monolayer, the C=O stretching band decreases drastically relatively to two bands at 1548 and 1390 $cm^{-1}$ that can be attributed to the asymmetric and symmetric stretching mode of the carboxylate function; on the spectrum of the $K^{W}_{Si}$[COOH] monolayer, the C=O vibration band decreases also and two carboxylate bands appear at 1570 and 1420 $cm^{-1}$.

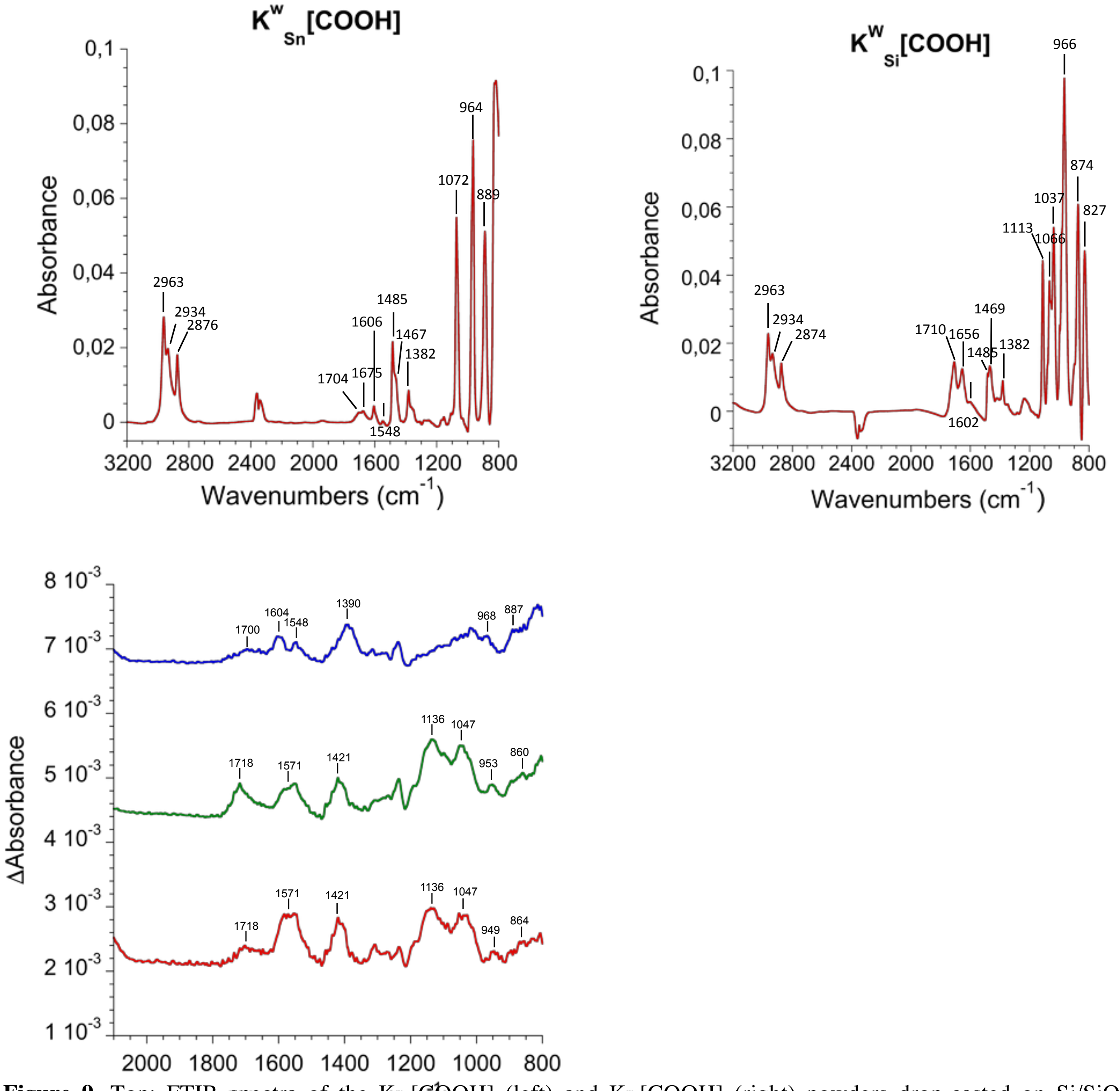

**Figure 9**. Top: FTIR spectra of the $K^{W}_{Sn}$[COOH] (left) and $K^{W}_{Si}$[COOH] (right) powders drop-casted on Si/$SiO_2$. Bottom: FTIR absorption spectra of the carboxylic acid POMs monolayers: $K^{W}_{Sn}$[COOH] layer (blue ; $K^{W}_{Si}$[COOH] layer (green ; $K^{W}_{Si}$[COOH] layer (red) after an added treatment by ultrasonication. The spectra are referenced to the substrate with its cleaned thermal oxide.

As in this case the C=O was still quite intense, an additional rinsing of the surface was performed in an ultrasonic bath. Interestingly, this new treatment induced a drastic decreasing of the C=O band on the FTIR spectrum and no modification of the carboxylate bands, showing that some additional physisorbed carboxylic acid POMs were removed by an additional rinsing, but that most of the POMs seem strongly attached to the surface by a carboxylate group. Moreover, as usually mentioned in carboxylate chemistry reports,[53,54] the separation between the carboxylate antisymmetric and symmetric stretching bands (Δν) can enlighten on the coordination mode of the carboxylate group: Δν=350-500 $cm^{-1}$ corresponds to a monodentate binding to the metal atom, Δν=150-180 $cm^{-1}$ corresponds to a bridging bidentate link to two adjacent metal atoms, and Δν=60-100 $cm^{-1}$ corresponds to a chelating bidentate binding mode. In the present case, the Δν separation is around 150 $cm^{-1}$ for both the $K^{W}_{Sn}$[COOH] and the $K^{W}_{Si}$[COOH] layers, which leads to a bidentate binding of the carboxylate to the substrate. However, discrimination between a bridging or a chelating mode is more awkward with a Δν value at the lower limit of the bridging bidentate mode for binding to two silicon atoms of the surface. As the C=O stretching band is still present, we cannot exclude that several anchoring modes are present on the surface and that some –COOH terminated POMs are attached via H-bonds, or that physisorbed species are still present on the surface. Note that in the case of the $K^{W}_{Si}$[COOH] layer, the typical longitudinal optical (LO) and transversal optical (TO) phonon modes of the silicon oxide appear at 1136 and 1047 $cm^{-1}$,[55] showing that the layer of $SiO_2$ slightly increased during the $K^{W}_{Si}$[COOH] deposition, probably due to the more disordered aspect of the $K^{W}_{Si}$[COOH] monolayer, which let place to the formation of random $SiO_2$ islands. The characteristic bands of the POMs below 1000 $cm^{-1}$, typically around 960 and 870 $cm^{-1}$ are yet detectable.

## Electrical measurements

Figure 10 shows the J-V curves in a semi-log plot measured with the mercury drop technique (see experimental section) on the silicon surface without POM deposition (Si/$SiO_2$ reference sample cut from the same wafer), compared with $K^{W}_{Sn}$[COOH] and $K^{W}_{Si}$[COOH] monolayer junctions. We clearly observe three distinct families of J-V curves. Typical current density histograms at 1V are also shown and were fitted by a log-normal distribution, with a log-mean of current density log-μ = -0.96, -1.53 and -3.23 for the Si/$SiO_2$ reference electrode, the $K^{W}_{Sn}$[COOH] and the $K^{W}_{Si}$[COOH] POM junctions, respectively (*i.e.* a mean current density of $1.1 \times 10^{-1}$, $3.0 \times 10^{-2}$ and $5.9 \times 10^{-4}$ A.cm$^{-2}$, respectively). We note that the dispersion of the current density is larger for the POMs samples than for the Si/$SiO_2$ sample (log standard deviations are log-σ = 0.3, 0.54 and 0.45 for the Si/$SiO_2$ reference electrode, the $K^{W}_{Sn}$[COOH] and $K^{W}_{Si}$[COOH] POM junctions, respectively), which may be related to the larger distribution of the POM monolayer thickness compared to the native $SiO_2$ (as assessed by ellipsometry and roughness AFM measurements).

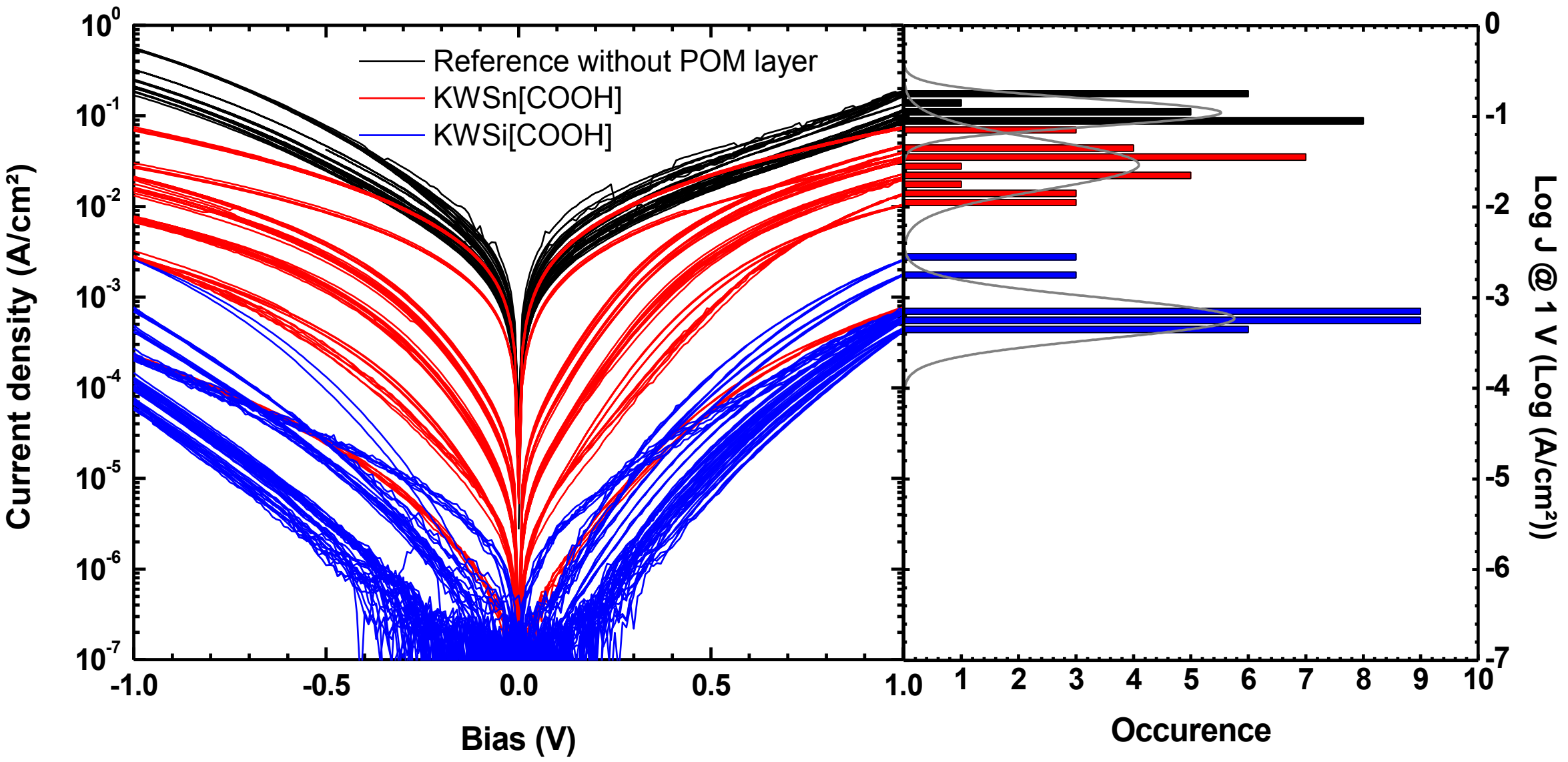

**Figure 10.** J–V curves measured for Si/$SiO_2$ reference substrate, $K^{W}_{Sn}$[COOH] and $K^{W}_{Si}$[COOH] monolayers on Si/$SiO_2$ (about 30 J-V curves, see details on the experimental section). Voltages were applied on the Hg drop and Si substrate was grounded.

As in previous works on similar POMs,[56] the experimental J-V curves are analyzed using an electron tunneling model. We used the so-called "modified Simmons model",[57-61] which introduces in the usual Simmons model for tunneling through a rectangular energy barrier[62] a unitless adjustable non-ideality "shape factor" to take into account: (i) the deviation from the simple rectangular barrier in molecular junctions and/or a more complex barrier (here, a multilayer structure composed of the silicon oxide, the organic tether and the POM) (ii) a poorly defined effective mass in molecular junction. The case α = 1 corresponds to an ideal rectangular barrier and a bare electron mass (reduced effective electron mass=1).

$$J = \left(\frac{e}{4\pi^2\hbar d^2}\right)\left\{\left(\Phi - \frac{eV}{2}\right)\times exp\left[-\frac{2(2m)^{1/2}}{\hbar}\alpha\left(\Phi - \frac{eV}{2}\right)^{1/2} d\right] - \left(\Phi + \frac{eV}{2}\right)\times exp\left[-\frac{2(2m)^{1/2}}{\hbar}\alpha\left(\Phi + \frac{eV}{2}\right)^{1/2} d\right]\right\} \qquad (1)$$

where J is the current density, e the electron charge, m is the bare electron mass, $\hbar$ the reduced Planck constant, d is the barrier width corresponding to the thickness of the layers (native oxide, 1.4 nm, plus POM monolayer) measured by ellipsometry, Φ is an effective barrier height, V is the applied bias and α the shape factor described above.

All I-V curves (figure 10) were fitted by Eq. 1 with the two parameters Φ and α (see typical fits in the supporting information, Figure S19). The histograms of the effective energy barrier (Φ) for the Si/$SiO_2$ sample, $K^{W}_{Sn}$[COOH] and $K^{W}_{Si}$[COOH] junctions are given in Figure 11 and fitted by a Gaussian law. We observe a clear offset towards higher values for the $K^{W}_{Sn}$[COOH] compared to the $K^{W}_{Si}$[COOH], with mean values of Φ= 1.75 eV (standard deviation 0.12 eV) and Φ= 1.39 eV (standard deviation 0.23 eV), respectively. For the reference sample, we have Φ= 1.91 eV (standard deviation 0.13 eV), which is consistent with previous values for native $SiO_2$.[63] We note that it seems counterintuitive to observe a higher current (figure 10) for the sample $K^{W}_{Sn}$[COOH] with a higher energy barrier (figure 11) and the thicker monolayer (see figures 2 and 3). This feature is understood if we consider the smaller values of the shape factor α (0.26-0.43) for this $K^{W}_{Sn}$[COOH] sample compared to α = 0.69-0.87 for the $K^{W}_{Si}$[COOH] sample (see supporting information Figure S20). In Eq. 1 the dominant term is exp(- $\alpha\Phi^{1/2}d$) and we have a smaller average value $\alpha\Phi^{1/2}d \approx 2.3$ for the $K^{W}_{Sn}$[COOH] sample than $\alpha\Phi^{1/2}d \approx 3.1$ for the $K^{W}_{Si}$[COOH] sample, thus consistent with a higher current for the former one (with average thicknesses d=5.1 nm and 3.3 nm, respectively, i.e. 1.4 nm for the native oxide plus the POM layer thickness, see ellipsometry results). We suggest that this difference of the shape factor is likely related to the nature of the organic tether of the POM hybrid, π-conjugated vs. σ-saturated. It is known that the exact potential distribution in a molecular junction is more complicated in a π-conjugated moiety, while it is more linear through alkyl chains (see a typical example in reference 62),[64] which is consistent with a shape factor closer to ideality in this latter case. Note also that we cannot exclude an additional effect of the $Pd^{2+}$ cations present in the $K^{W}_{Sn}$[COOH] layers on the intensity of the observed current.

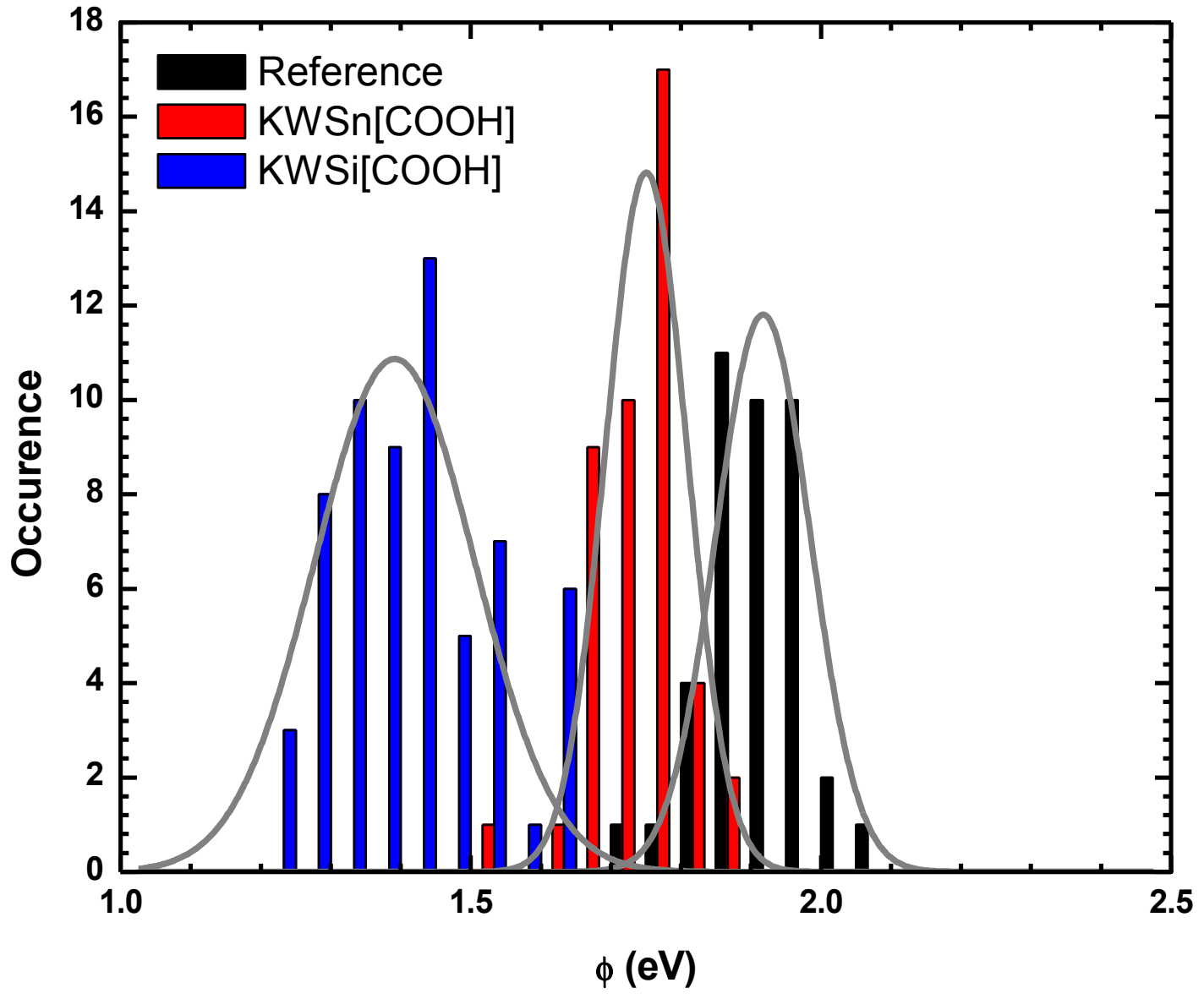

**Figure 11.** Effective barrier height Φ histograms for $K^{W}_{Sn}$[COOH], $K^{W}_{Si}$[COOH] monolayers and the reference bare Si/$SiO_2$ sample.

We can further refine the electronic structure analysis of the Si/$SiO_2$/POM junctions. The effective energy barrier is made of two parts, the native $SiO_2$ (Φ ≈ 1.9 eV, d ≈ 1.4 nm) and the POM monolayer characterized by the energy position of the POM LUMO, $\varepsilon_{POM}$, with respect to the Fermi energy of the electrodes. We ascribe the molecular orbital involved in the electron transport to the LUMO of the POMs, which are the closest to the Fermi energy of electrodes (see DFT calculations) while the HOMOs are deeper in energy. The lower Φ for the two POM samples means that the LUMO is lower in energy. Assuming a simple staircase energy barrier model (Figure 12 , details in the supporting information and reference 63),[65] we can deduce the energy position of the POM LUMO, $\varepsilon_{POM}$ at 1.1 eV and 1.7 eV for the $K^{W}_{Si}$[COOH] and $K^{W}_{Sn}$[COOH] samples, respectively. Finally, we compare these LUMO energy positions, with the ones determined from cyclic voltammograms and DFT calculations (Figure 12). The LUMO positions from CV in solution and the DFT calculations (continuum solvation model) are in good quantitative agreement, and in good qualitative agreement with the values from the electrical measurements on the "solid-state" molecular device (LUMO of $K^{W}_{Si}$[COOH] lower than $K^{W}_{Sn}$[COOH]). The LUMO energy differences are consistent for the three methods (0.6 eV). The fact that the LUMOs are shifted in the molecular junctions may be due to molecule/electrode coupling and charge transfer and/or the presence of trapped charges in the native $SiO_2$ layer. In solution, solvation is also expected to stabilize the energy levels.

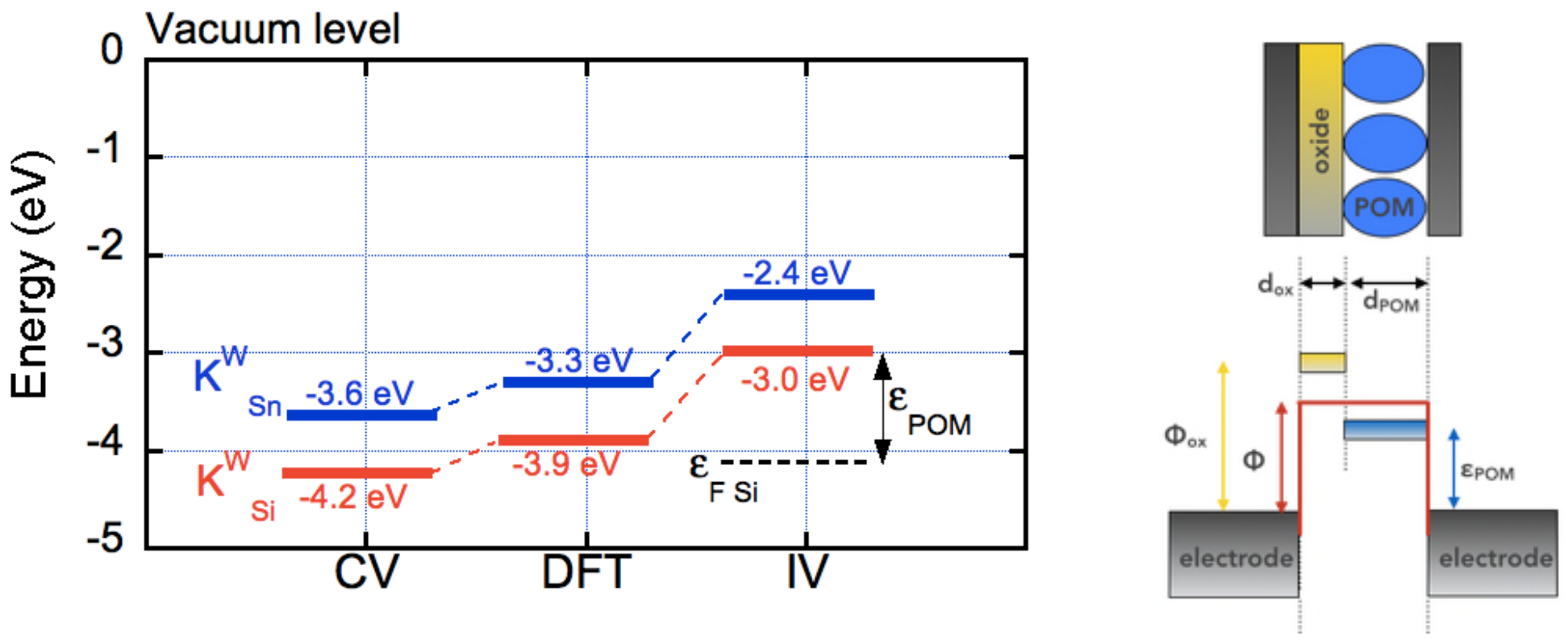

**Figure 12.** LUMO energy positions with respect to the vacuum level. The DFT position is given by the first empty d orbital of the POMs (see "DFT section"). The CV position is calculated from the first reduction waves (Figure S10), $E_{LUMO}=-(E_{1/2} + E_{SCE/SHE})-4.4$ (in eV with $E_{SCE/SHE}=0.24$ eV).[66] The IV position is obtained from the $\varepsilon_{POM}$ value considering the $n^{+}$ doped silicon Fermi energy at -4.1 eV (Si electron affinity) and the junction energy diagram shown in the right (see details in the Supporting Information and in a previous publication[65]).

Finally, we note that the LUMO position of the $K^{W}_{Sn}$[COOH] POM is consistent with the position of similar molecule $K^{W}_{Sn}[N_2^{+}]$ grafted on hydrogen-terminated silicon (no oxide) using a diazonium group instead of a carboxylic acid one ($\varepsilon_{POM}$ = 1.8 eV in reference 53 vs. 1.7 eV here),[56] which indicates that the electronic structure of these molecular junctions is mainly controlled by the POM LUMO. As noted above the LUMOs of the POM hybrids are combination of tungsten d orbitals and do not involve the orbitals of the tether or the anchor group. The shifting down of about 0.6 eV on going from $K^{W}_{Sn}$[COOH] to $K^{W}_{Si}$[COOH] (molecule in solution, in a solvation continuum model, as well as for the "solid-state" self-assembled monolayers) is rather attributable to the change in the total charge (4- for $K^{W}_{Sn}$[COOH] *vs* 3- for $K^{W}_{Si}$[COOH]).[67]

## CONCLUSION

Two POM hybrids $TBA_{4.4}[PW_{11}O_{39}\{Sn(C_6H_4)C{\equiv}C(C_6H_4)COOH_{0.6}\}]$ ($K^{W}_{Sn}$[COOH]) and $TBA_{3.4}[PW_{11}O_{39}\{O(SiC_2H_4COOH_{0.8})_2\}]$ ($K^{W}_{Si}$[COOH]) have been prepared and thoroughly characterized in solution by NMR, UV-visible spectroscopies and cyclic-voltammetry. They display remote carboxylic acid/carboxylate groups that allow investigating their covalent grafting onto Si/$SiO_2$ flat substrates. Monolayers have been obtained and characterized by ellipsometry, AFM, and XPS. The $K^{W}_{Si}$[COOH] layer was found to be less uniform than the $K^{W}_{Sn}$[COOH] layer, presumably because of the presence of two, moreover more flexible, aliphatic arms. An FTIR study was carried out on a special low doped float-zone silicon substrate to probe the POM binding mode, from which a bidentate carboxylate was inferred. Finally, the electronic properties of the POM hybrids disclosed by cyclic-voltammetry, electronic absorption spectroscopy and calculated energy level diagrams have been confronted to the energetics of the Si/$SiO_2$//POM layer//Hg junction as probed by electron transport measurements at the solid state.

These three approaches are consistent, showing that the energy position of the LUMO of $K^{W}_{Si}$[COOH] is lower than the one of $K^{W}_{Sn}$[COOH] by a difference of 0.6 eV, in agreement with a higher total charge for the later. The LUMO energy of $K^{W}_{Sn}$[COOH] grafted on Si/$SiO_2$ and of $K^{W}_{Sn}[N_2^+]$ grafted on silicon (no oxide) are similar, showing that the electronic properties of these molecular junctions are driven by the POM LUMO (d-orbitals) and not by the π-orbitals of the tether or the anchor group. Such POM-based electrodes represent new functionalized silicon-based electrodes, which have important implications for improving and/or tuning the electronic properties of a variety of electronic devices: photovoltaics, sensors, memories and bioelectronics.[68] Some of us have recently shown that the controlled grafting of POMs on various oxides is also relevant to energy conversion.[69]

ASSOCIATED CONTENT

**Supporting Information** file includes: $^{1}H$ and $^{31}P$ NMR spectra of $K^{W}_{Sn}$[COOH] and $K^{W}_{Si}$[COOH]; Infrared spectrum of $K^{W}_{Sn}$[COOH] and $K^{W}_{Si}$[COOH] in KBr pellets; ESI$^{-}$ spectra of 1 µmol.L$^{-1}$ of $K^{W}_{Sn}$[COOH] and $K^{W}_{Si}$[COOH] in acetonitrile ; cyclic voltammograms of $K^{W}_{Sn}$[COOH] in acetonitrile; thickness histograms inferred from ellipsometry; AFM images of the $K^{W}_{Sn}$[COOH] and $K^{W}_{Si}$[COOH] layers; XPS characterization; FTIR spectra of the $K^{W}_{Sn}$[COOH] and $K^{W}_{Si}$[COOH] powders drop-casted on Si/$SiO_2$; example of fits of IV curves using the modified Simmons model; description of the two layers staircase energy model; XYZ coordinates from the DFT calculation. The file is available free of charge as a PDF file.

AUTHOR INFORMATION

**Corresponding Author**

**Anna Proust** - *Sorbonne Université, CNRS, Institut Parisien de Chimie Moléculaire, IPCM, 4 Place Jussieu, F-75005 Paris, France*; orcid.org/0000-0002-0903-6507;

Email : anna.proust@sorbonne-universite.fr

**Authors**

**Maxime Laurans, Kelly Trinh, Kevin Dalla Francesca, Guillaume Izzet, Sandra Alves, Etienne Derat, Florence Volatron** - *Sorbonne Université, CNRS, Institut Parisien de Chimie Moléculaire, IPCM, 4 Place Jussieu, F-75005 Paris, France*

**Kelly Trinh, Olivier Pluchery** - *Sorbonne Université, CNRS, Institut des Nanosciences de Paris, INSP, 4 Place Jussieu, F-75005 Paris, France*

**Dominique Vuillaume, Stéphane Lenfant** - *Institute for Electronics Microelectronics and Nanotechnology (IEMN), CNRS, Av. Poincaré, Villeneuve d'Ascq, France*

**Vincent Humblot** - *Sorbonne Université, CNRS, Laboratoire de réactivité de surface, LRS, 4 Place Jussieu, F-75005 Paris, France- present address : FEMTO-ST Institute, UMR CNRS 6174, Université Bourgogne Franche-Comté, 15B avenue des Montboucons, 25030 Besançon Cedex, France*

## Author Contributions

The manuscript was written through contributions of all authors. All authors have given approval to the final version of the manuscript.

## Funding Sources

This work was supported by Sorbonne Université and by the CNRS. K.D.F. thanks the program PER-SU of Sorbonne Universités for his post-doctoral grant.

ACKNOWLEDGMENT

The authors acknowledge IMPC from Sorbonne University (Institut des Matériaux de Paris Centre, FR CNRS 2482) and the C'Nano projects of the Region Ile-de-France, for Omicron XPS apparatus funding.

ABBREVIATIONS

POM, polyoxometalate; AFM, atomic force microscopy; XPS, X-ray photoelectron spectroscopy; FTIR, Fourier Transform Infra Red.

SYNOPSIS

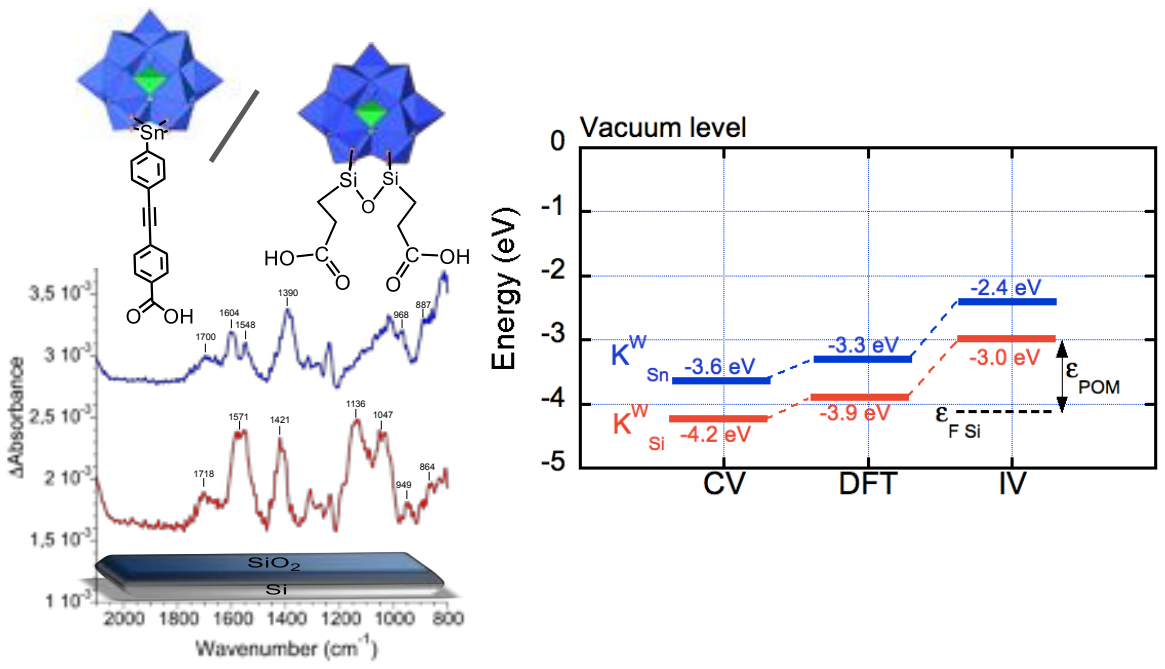

# SUPPORTING INFORMATION

# Covalent Grafting of Polyoxometalate Hybrids onto Flat Silicon/Silicon Oxide: Insights from POMs Layers on Oxides

*Maxime Laurans,[a] Kelly Trinh,[a,d] Kevin Dalla Francesca,[a] Guillaume Izzet,[a] Sandra Alves,[a] Etienne Derat,[a] Vincent Humblot,[c,†] Olivier Pluchery,[d] Dominique Vuillaume,[b] Stéphane Lenfant,[b] Florence Volatron,[a] Anna Proust*.[a]*

[a]Sorbonne Université, CNRS, Institut Parisien de Chimie Moléculaire, IPCM, 4 Place Jussieu, F-75005 Paris, France ; anna.proust@sorbonne-universite.fr

[b] Institute for Electronics Microelectronics and Nanotechnology (IEMN), CNRS, Av. Poincaré, Villeneuve d'Ascq, France

[c] Sorbonne Université, CNRS, Laboratoire de réactivité de surface, LRS, 4 Place Jussieu, F-75005 Paris, France

[†] present address: FEMTO-ST Institute, UMR CNRS 6174, Université Bourgogne Franche-Comté, 15B avenue des Montboucons, 25030 Besançon Cedex, France

[d] Sorbonne Université, CNRS, Institut des Nanosciences de Paris, INSP, 4 Place Jussieu, F-75005 Paris, France

## 1. Characterization of $K^W_{Sn}$[COOH]

**Figure S1.** $^1$H (400 MHz) and $^{31}$P (121 MHz, frame inset) NMR spectra of $K^W_{Sn}$[COOH].

**Figure S2.** Infrared spectrum of $K^W_{Sn}$[COOH] in KBr pellet.

**Figure S3.** ESI$^-$ spectrum of 1 µmol.L$^{-1}$ of $K^W_{Sn}$[COOH] in ACN.

**Figure S4.** Comparison of experimental (upper traces) and calculated (lower traces) for $[PW_{11}O_{39}Sn(C_{14}H_8)COOH]^{4-}$ (top, left), $\{TBA[PW_{11}O_{39}Sn(C_{14}H_8)COOH]\}^{3-}$ (top, right), $[PW_{11}O_{39}Sn(C_{14}H_8)COO]^{5-}$ (bottom, left), $\{TBA_2[PW_{11}O_{39}Sn(C_{14}H_8)COOH]\}^{2-}$ (bottom, right), most abundant POM-based hybrid ions of $K^w_{Sn}$[COOH].

**Figure S5.** Cyclic voltammograms of $K^W_{Sn}$[COOH] in acetonitrile with $TBAPF_6$ 0.1 M at various scan rates ; 10 mV.s$^{-1}$ (black curve) ; 50 mV.s$^{-1}$ (blue curve) ; 100 mV.s$^{-1}$ (red curve) ; 200 mV.s$^{-1}$ (green curve) ; 300 mV.s$^{-1}$ (orange curve) ; 400 mV.s$^{-1}$ (purple curve).

## 2. Characterization of $K^W_{Si}$[COOH]

**Figure S6.** $^1$H (400 MHz) and $^{31}$P (121 MHz, frame inset) NMR spectra of $K^W_{Si}$[COOH].

**Figure S7.** Infrared spectrum of $K^W_{Si}$[COOH] in KBr pellet.

**Figure S8.** ESI$^-$ spectrum of 1 µmol.L$^{-1}$ of $K^W_{Si}$[COOH] in ACN.

**Figure S9.** Comparison of experimental (upper traces) and calculated (lower traces) for $[PW_{11}O_{39}Si\{O(SiC_2H_4COOH)_2\}]^{3-}$ (top, left), $<TBA[PW_{11}O_{39}\{O(SiC_2H_4COOH)(SiC_2H_4COO)\}]>^{3-}$ (top, right), $<TBA[PW_{11}O_{39}\{O(SiC_2H_4COOH)_2\}]>^{2-}$ (bottom, left), $<TBA_2[PW_{11}O_{39}\{O(SiC_2H_4COOH)(SiC_2H_4COO)\}]>^{2-}$ (bottom, right), most abundant POM-based hybrid ions of $K^w_{Si}$[COOH].

**Figure S10.** Cyclic voltammograms of $K^W_{Sn}$[COOH] ($E_1$= -1.01 V vs SCE) and $K^W_{Si}$[COOH] (($E''_1$= -0.38 V vs SCE) in acetonitrile with $TBAPF_6$ 0.1 M. Scan rate = 100 mV.s$^{-1}$.

## 3. Ellipsometry measurements

**Figure S11.** Thickness histograms for $K^W_{Sn}$[COOH]. Left: the total thickness as measured ($SiO_2$ layer plus POM layer), right: the $SiO_2$ thickness (1.4 nm) was subtracted to the total thickness values.

**Figure S12.** Thickness histograms for $K^W_{Si}$[COOH]. Left: the total thickness as measured ($SiO_2$ layer plus POM layer), right: the $SiO_2$ thickness (1.4 nm) was subtracted to the total thickness values.

## 4. AFM measurements

**Figure S13.** 5*5 µm AFM images and Z-profiles of (a) the bare Si/$SiO_2$ substrate, (b) the $K^W_{Sn}$[COOH] layer and (c) the $K^W_{Si}$[COOH] layer.

## 5. XPS measurements

**Figure S14.** N 1s, O 1s, Si 2p and P 2p high-resolution XPS spectra of the $K^{W}_{Sn}$[COOH] layer on $Si/SiO_2$.

**Figure S15.** Survey and Pd 3d high resolution spectra of the (a) $K^{W}_{Sn}$[COOH] powder and the (b) $K^{W}_{Sn}$[COOH] layer on $Si/SiO_2$.

**Figure S16.** Survey and W 4f, Pd 3d high-resolution XPS spectra of the $K^{W}_{Sn}$[H] layer on $Si/SiO_2$.

**Figure S17.** Survey, O 1s, Si 2p and Pd 3d high-resolution XPS spectra of the $K^{W}_{Si}$[COOH] layer on $Si/SiO_2$.

## 6. Fits of the modified Simmons model.

**Figure S18.** (left) Fit (red line) on a IV curve for $K^{W}_{Si}$[COOH] molecular junctions, with $\Phi$ = 1.3 eV and $\alpha$ = 0.82; (right) Fit (red line) on a IV curve for $K^{W}_{Sn}$[COOH] molecular junctions, with $\Phi$ = 1.7 eV and $\alpha$ = 0.37.

**Figure S19.** $\alpha$ parameter histograms for $K^{W}_{Sn}$[COOH], $K^{W}_{Si}$[COOH] monolayers and the reference bare $Si/SiO_2$ sample. The mean of $\alpha$ parameter $\alpha$ = 1.21 ± 0.06, 0.77 ± 0.13 and 0.36 ± 0.08 and for the $Si/SiO_2$ reference electrode, the $K^{W}_{Si}$[COOH] and the $K^{W}_{Sn}$[COOH] POM junctions, respectively.

## 7. Two layers staircase energy model

## 8. XYZ coordinates from DFT calculations

## 1. Characterization of $K^{W}_{Sn}[COOH]$

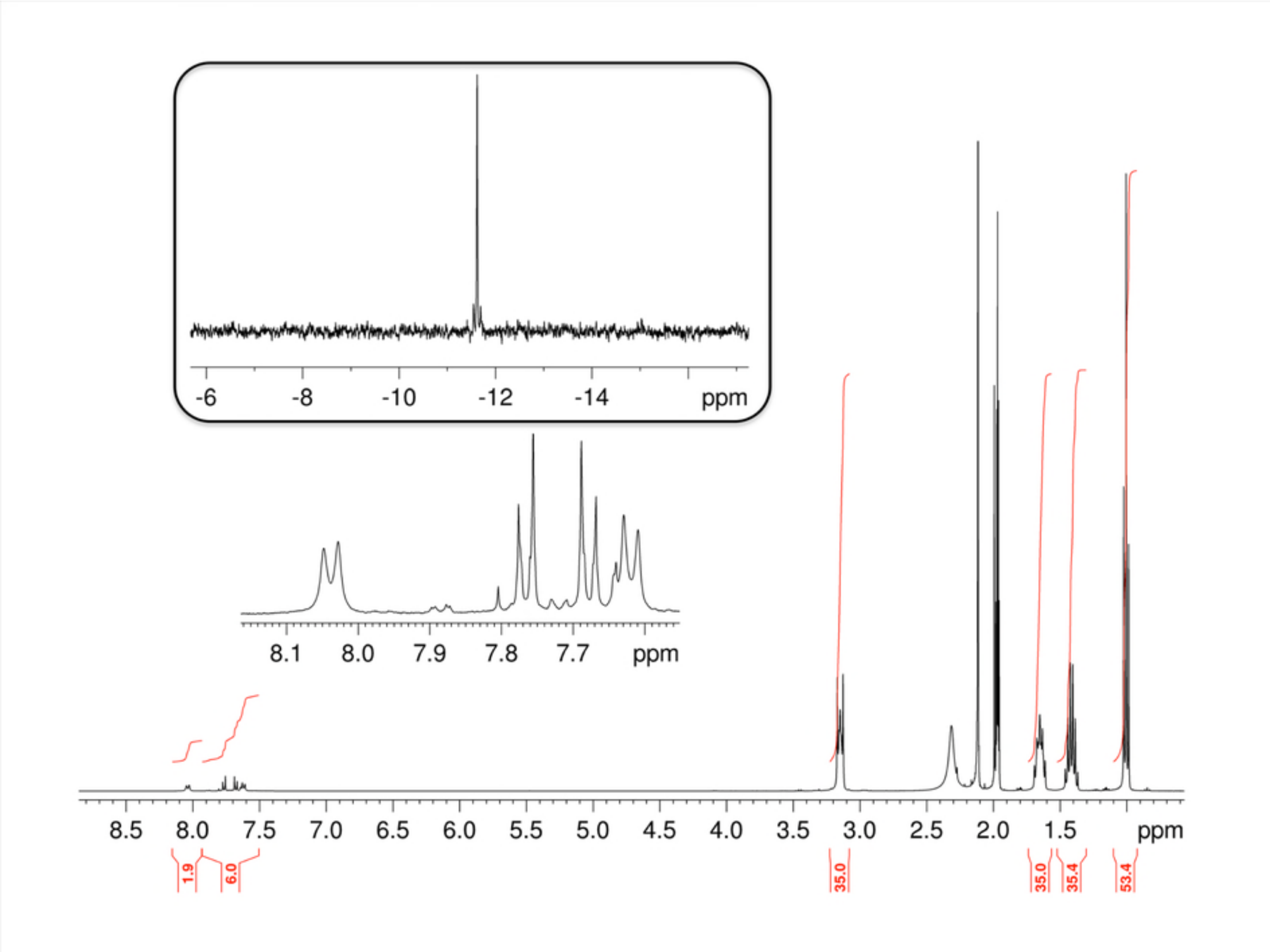

**Figure S1.** $^{1}H$ (400 MHz) and $^{31}P$ (121 MHz, frame inset) NMR spectra of $K^{W}_{Sn}[COOH]$.

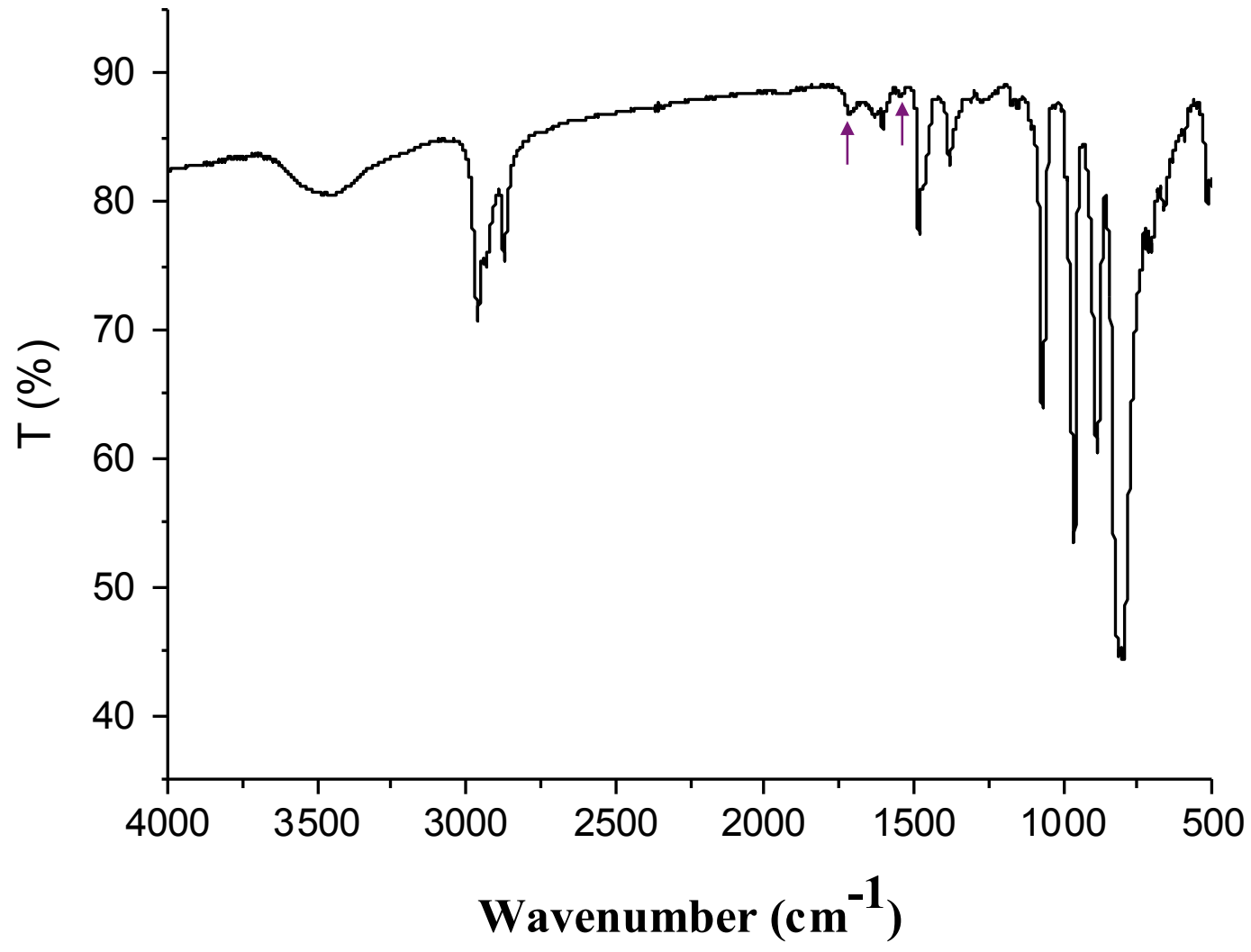

**Figure S2.** Infrared spectrum of $K^{W}_{Sn}[COOH]$ in KBr pellet.

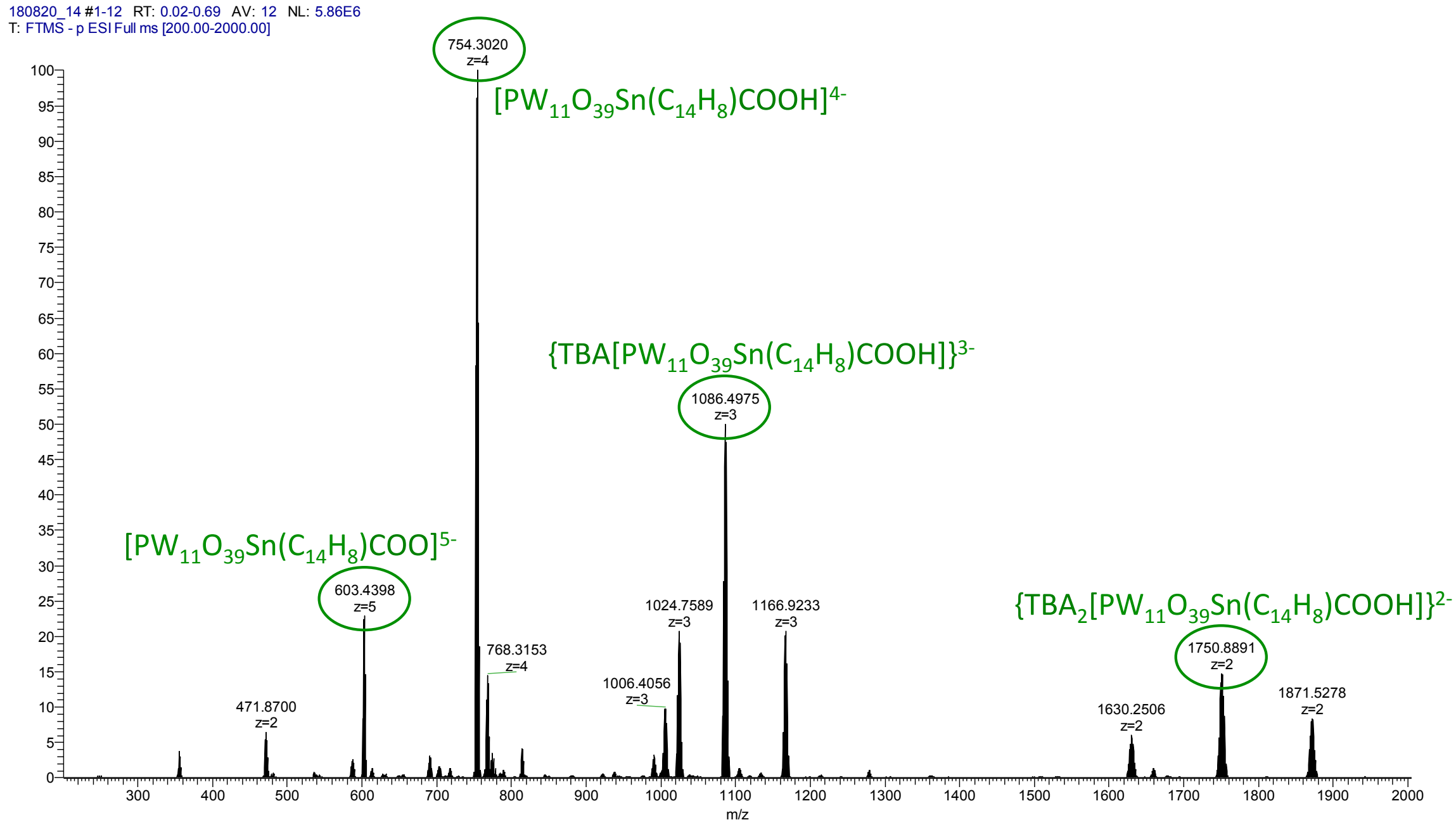

**Figure S3.** ESI$^-$ spectrum of 1 μmol.L$^{-1}$ of $K^W{}_{Sn}$[COOH] in ACN.

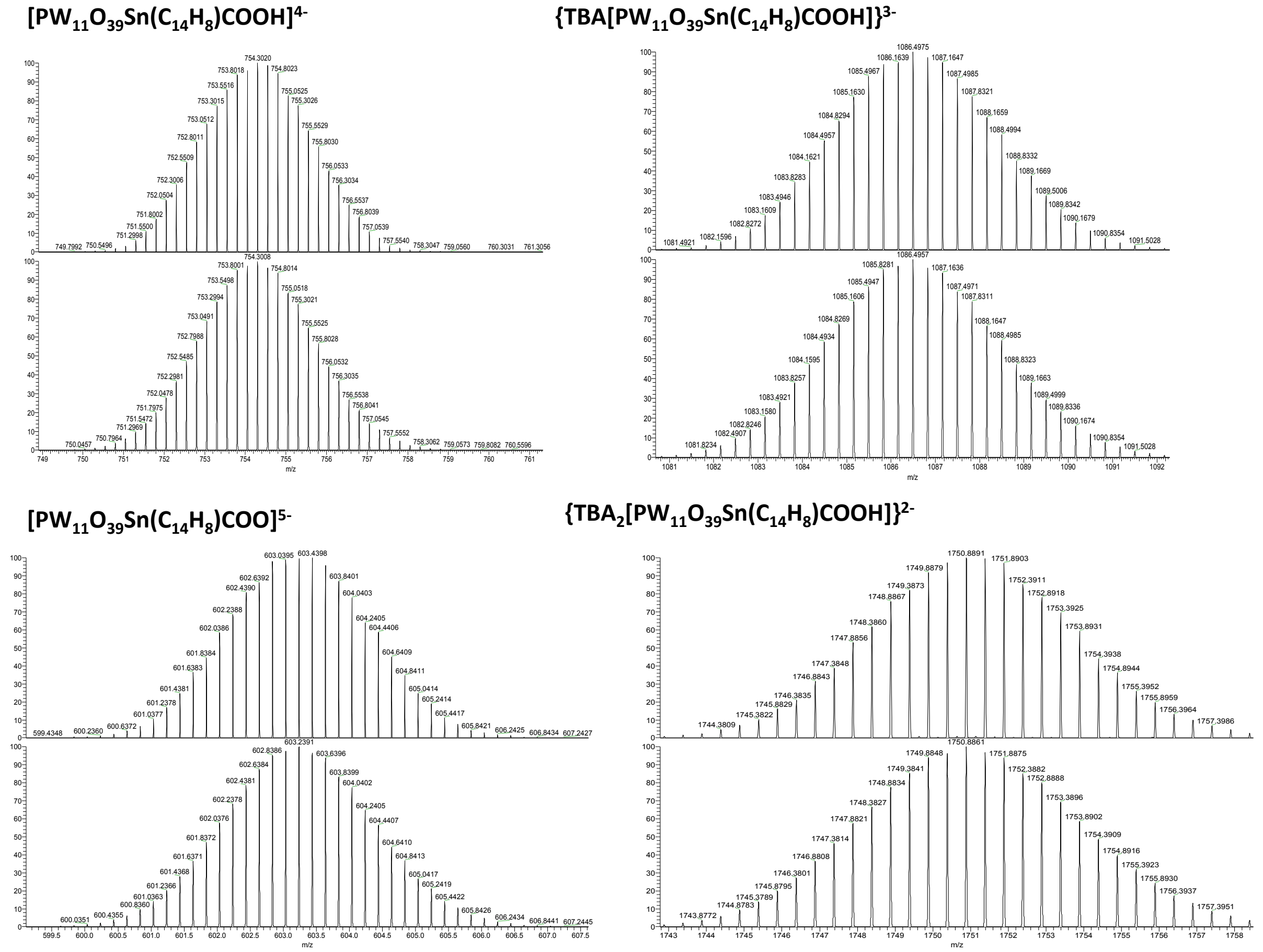

**Figure S4.** Comparison of experimental (upper traces) and calculated (lower traces) for $[PW_{11}O_{39}Sn(C_{14}H_8)COOH]^{4-}$ (top, left), $\{TBA[PW_{11}O_{39}Sn(C_{14}H_8)COOH]\}^{3-}$ (top, right), $[PW_{11}O_{39}Sn(C_{14}H_8)COO]^{5-}$ (bottom, left), $\{TBA_2[PW_{11}O_{39}Sn(C_{14}H_8)COOH]\}^{2-}$ (bottom, right), most abundant POM-based hybrid ions of $K^w{}_{Sn}$[COOH].

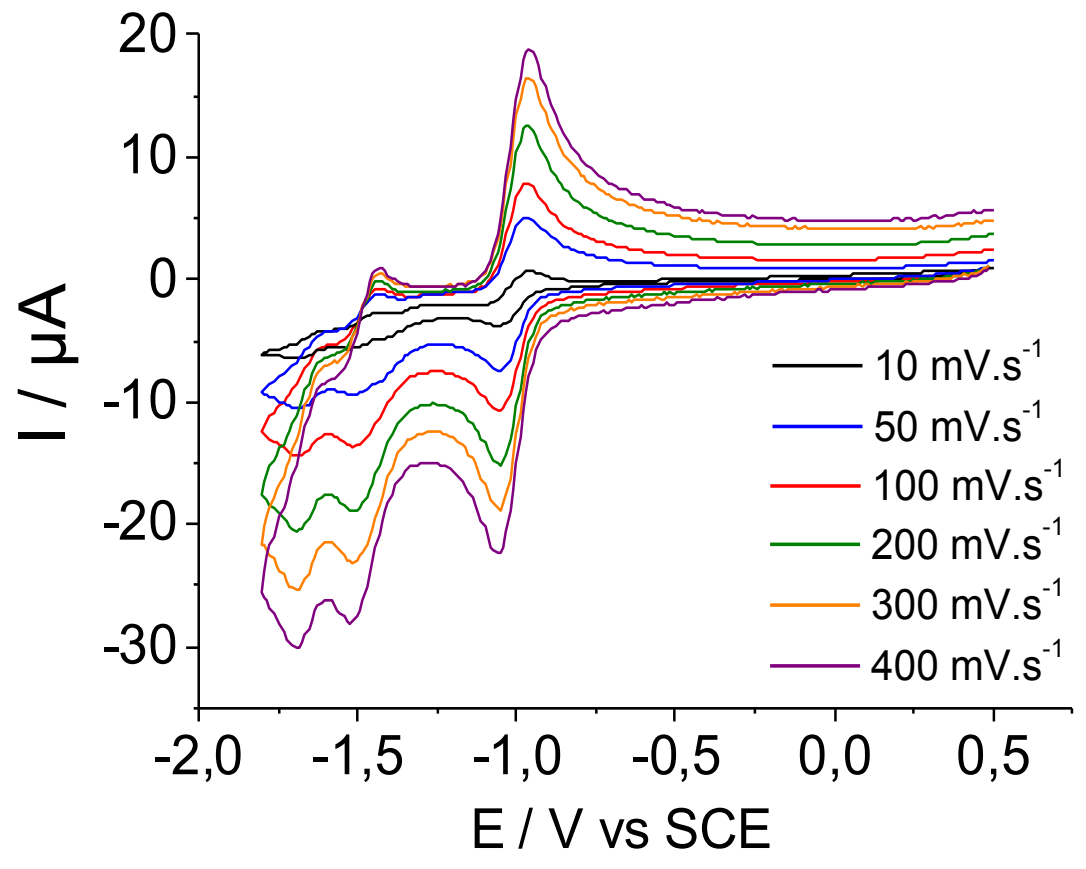

**Figure S5.** Cyclic voltammograms of $K^{W}_{Sn}$[COOH] in acetonitrile with $TBAPF_6$ 0.1 M at various scan rates ; 10 mV.s$^{-1}$ (black curve) ; 50 mV.s$^{-1}$ (blue curve) ; 100 mV.s$^{-1}$ (red curve) ; 200 mV.s$^{-1}$ (green curve) ; 300 mV.s$^{-1}$ (orange curve) ; 400 mV.s$^{-1}$ (purple curve).

## 2. Characterization of $K^{W}_{Si}$[COOH]

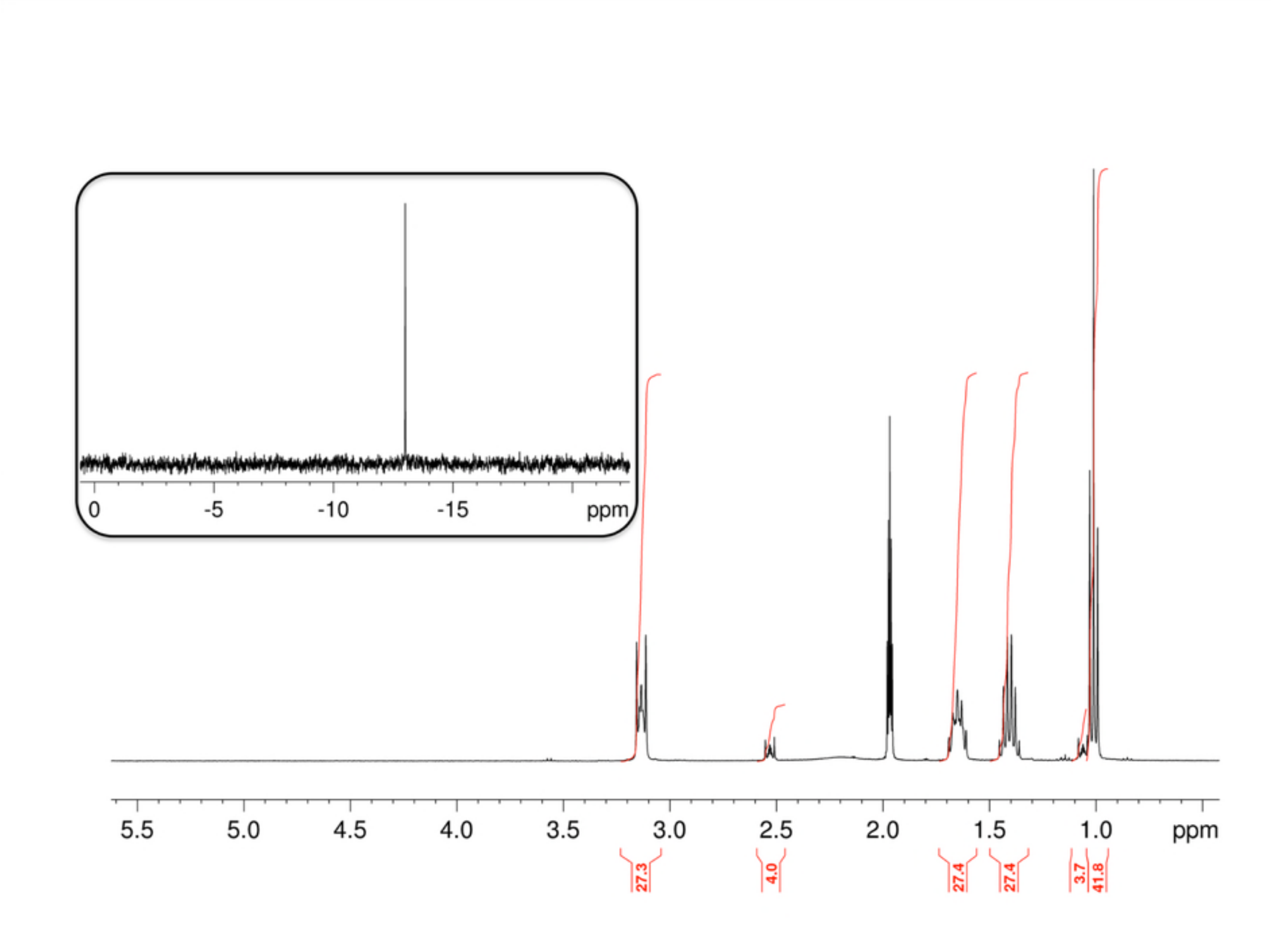

**Figure S6.** $^{1}$H (400 MHz) and $^{31}$P (121 MHz, frame inset) NMR spectra of $K^{W}_{Si}$[COOH].

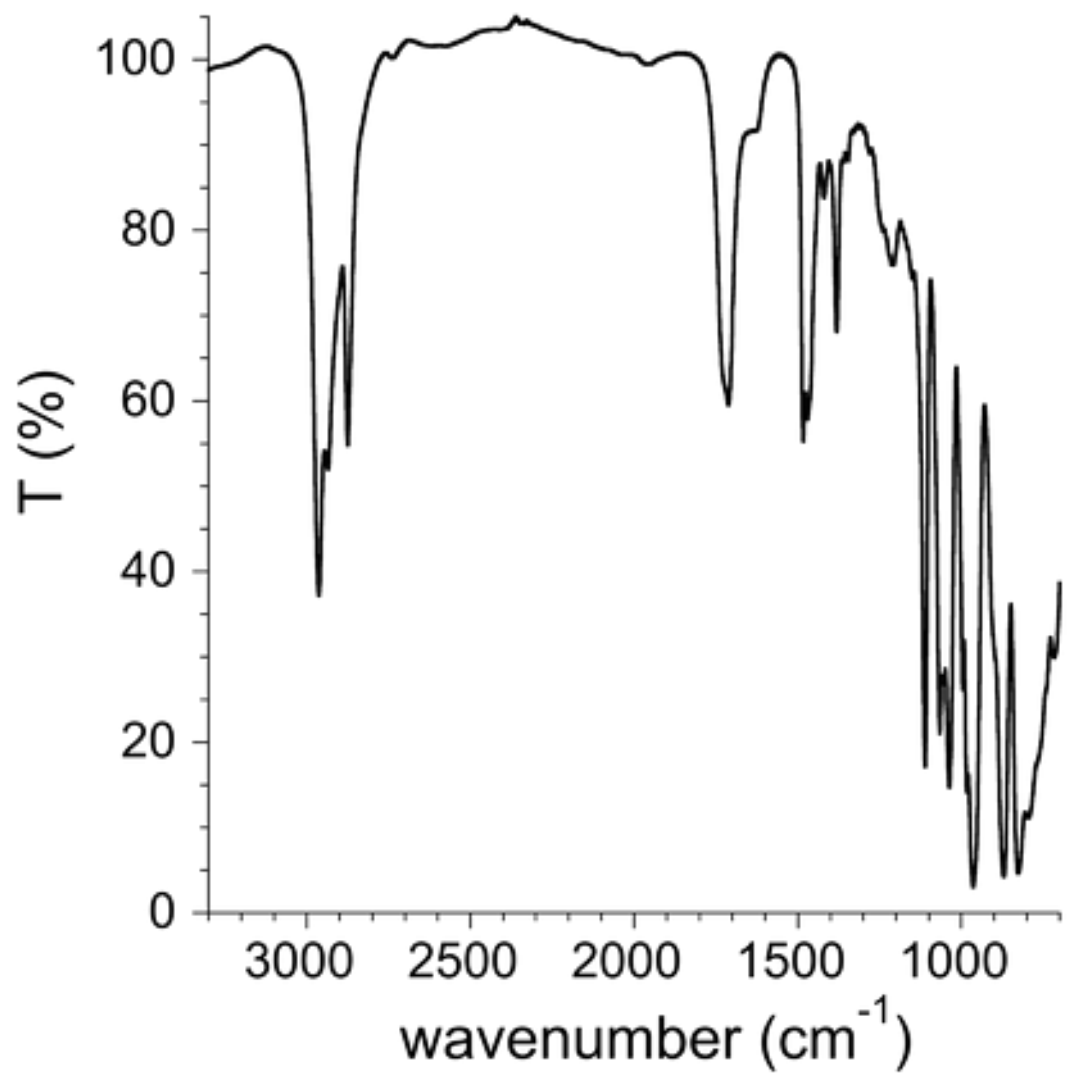

**Figure S7.** Infrared spectrum of $K^{W}_{Si}$[COOH] in KBr.

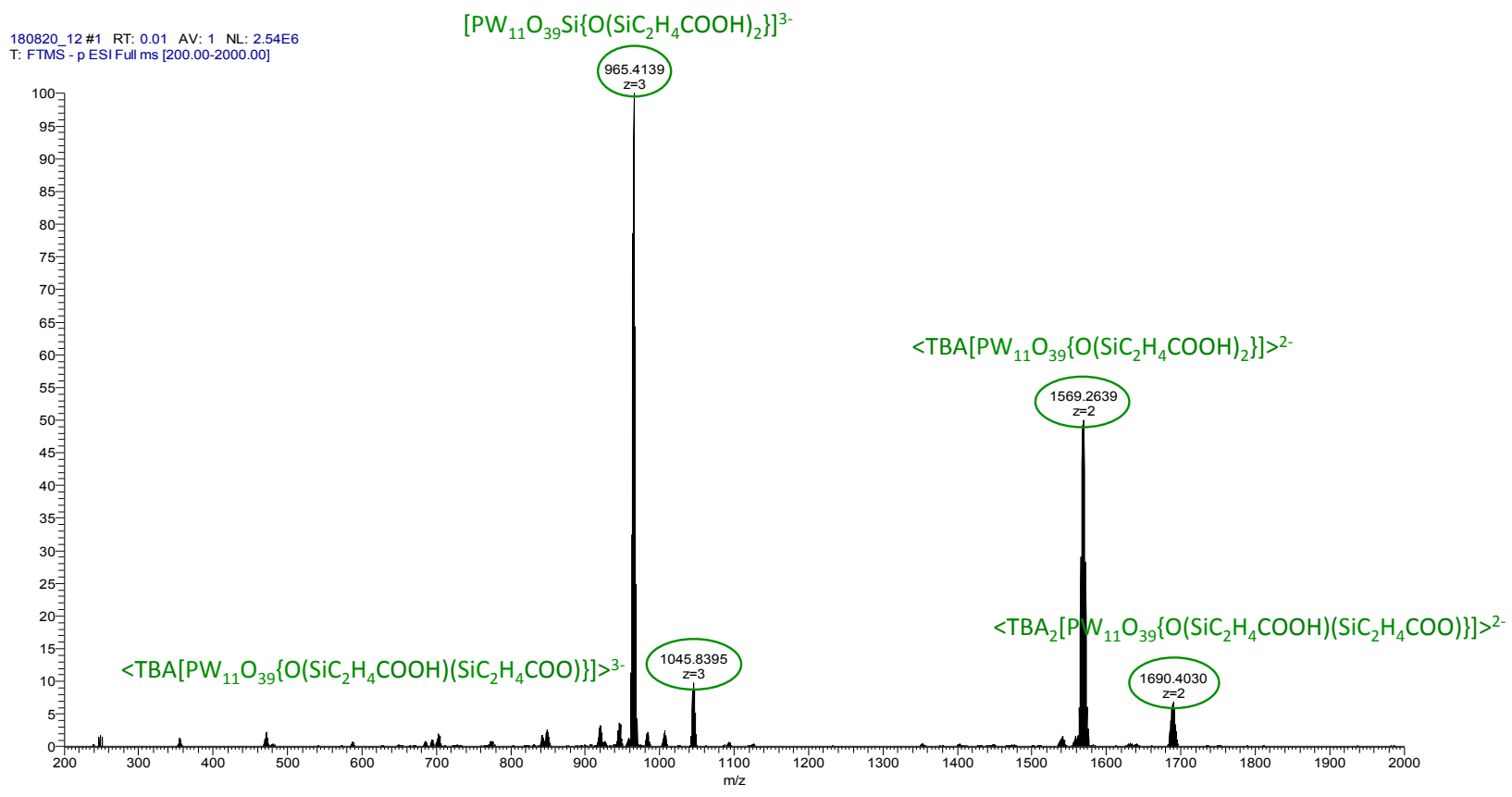

**Figure S8.** ESI$^-$ spectrum of 1 µmol.L$^{-1}$ of $K^{W}_{Si}$[COOH] in ACN.

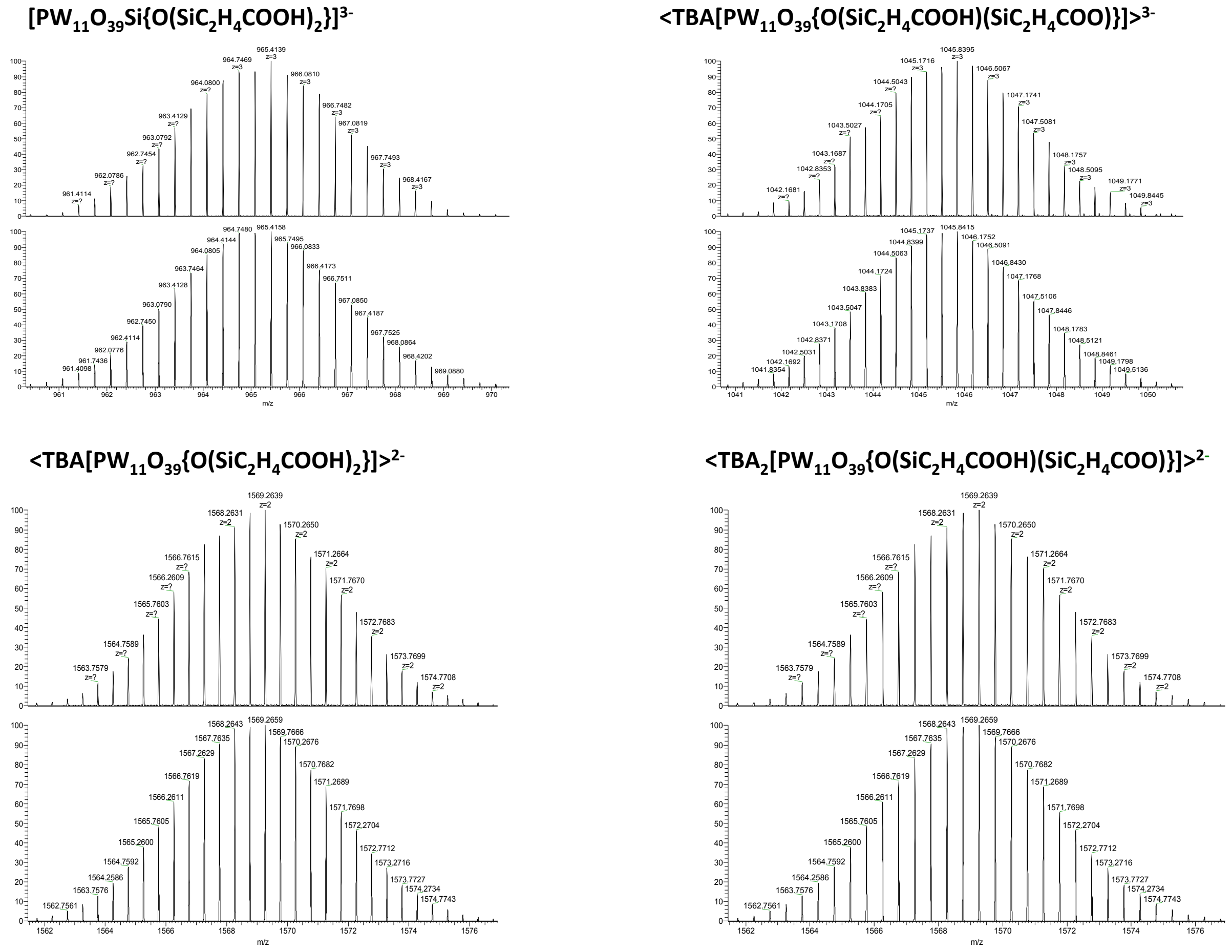

**Figure S9.** Comparison of experimental (upper traces) and calculated (lower traces) for $[PW_{11}O_{39}Si\{O(SiC_2H_4COOH)_2\}]^{3-}$ (top, left), <$TBA[PW_{11}O_{39}\{O(SiC_2H_4COOH)(SiC_2H_4COO)\}]$>$^{3-}$ (top, right), <$TBA[PW_{11}O_{39}\{O(SiC_2H_4COOH)_2\}]$>$^{2-}$ (bottom, left), <$TBA_2[PW_{11}O_{39}\{O(SiC_2H_4COOH)(SiC_2H_4COO)\}]$>$^{2-}$ (bottom, right), most abundant POM-based hybrid ions of $K^{w}_{Si}$[COOH].

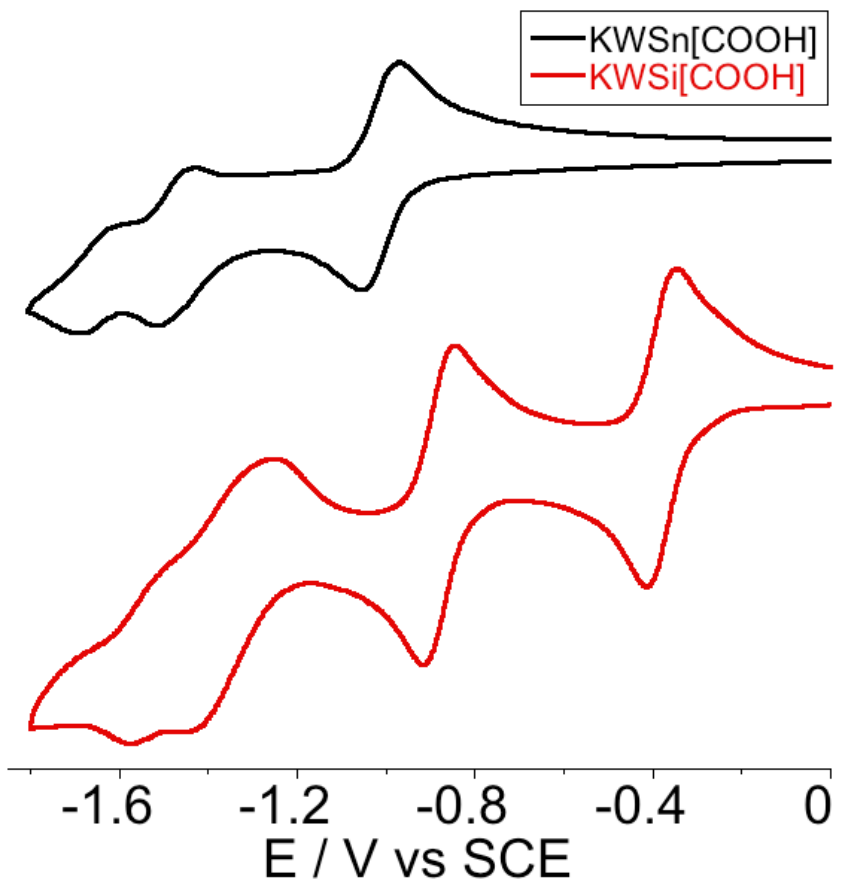

**Figure S10.** Cyclic voltammograms of $K^{W}_{Sn}$[COOH] ($E_1$= -1.01 V vs SCE) and $K^{W}_{Si}$[COOH] (($E'_1$= -0.38 V vs SCE) in acetonitrile with $TBAPF_6$ 0.1 M. Scan rate = 100 mV.$s^{-1}$.

## 3. Ellipsometry measurements

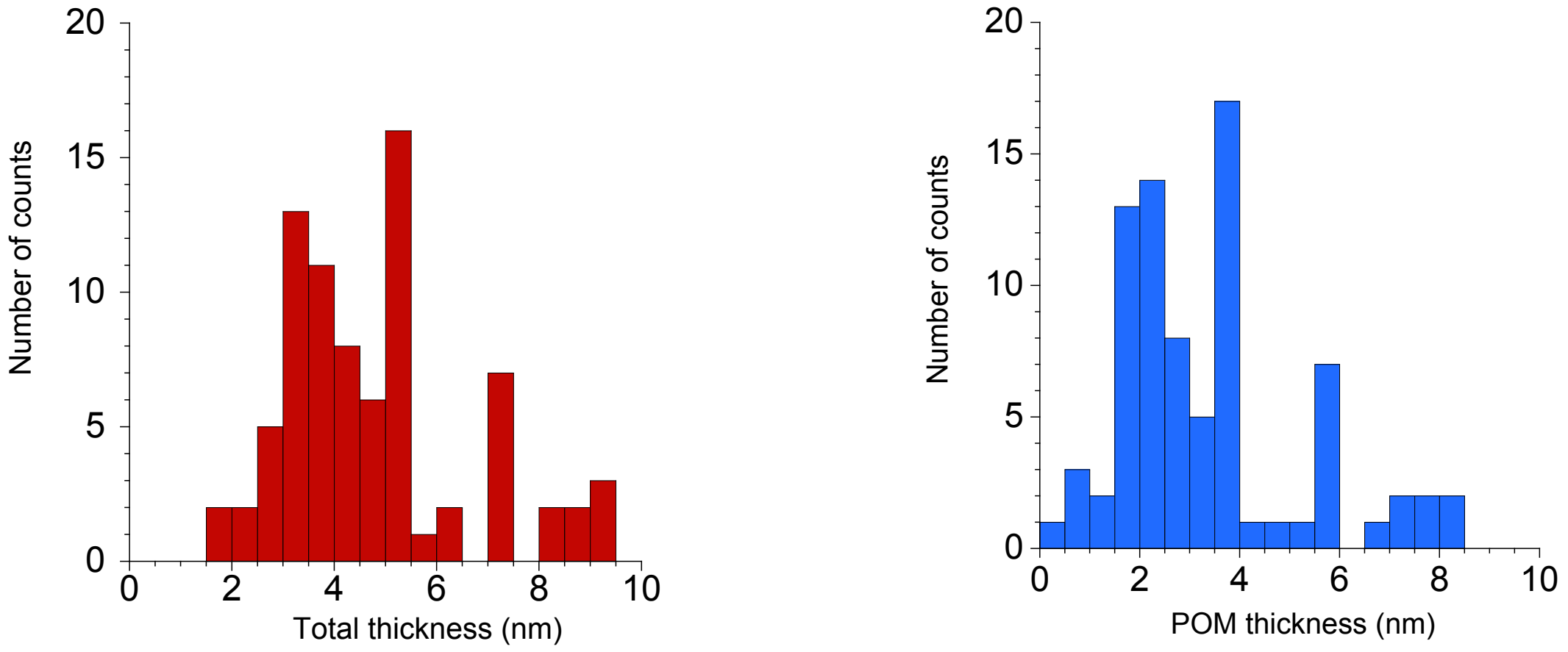

**Figure S11.** Thickness histograms for $K^{W}_{Sn}$[COOH]. Left: the total thickness as measured ($SiO_2$ layer plus POM layer); right: the $SiO_2$ thickness (1.4 nm) was subtracted to the total thickness values. Most of the $K^{W}_{Sn}$[COOH] thickness values range from around 2 to 4 nm.

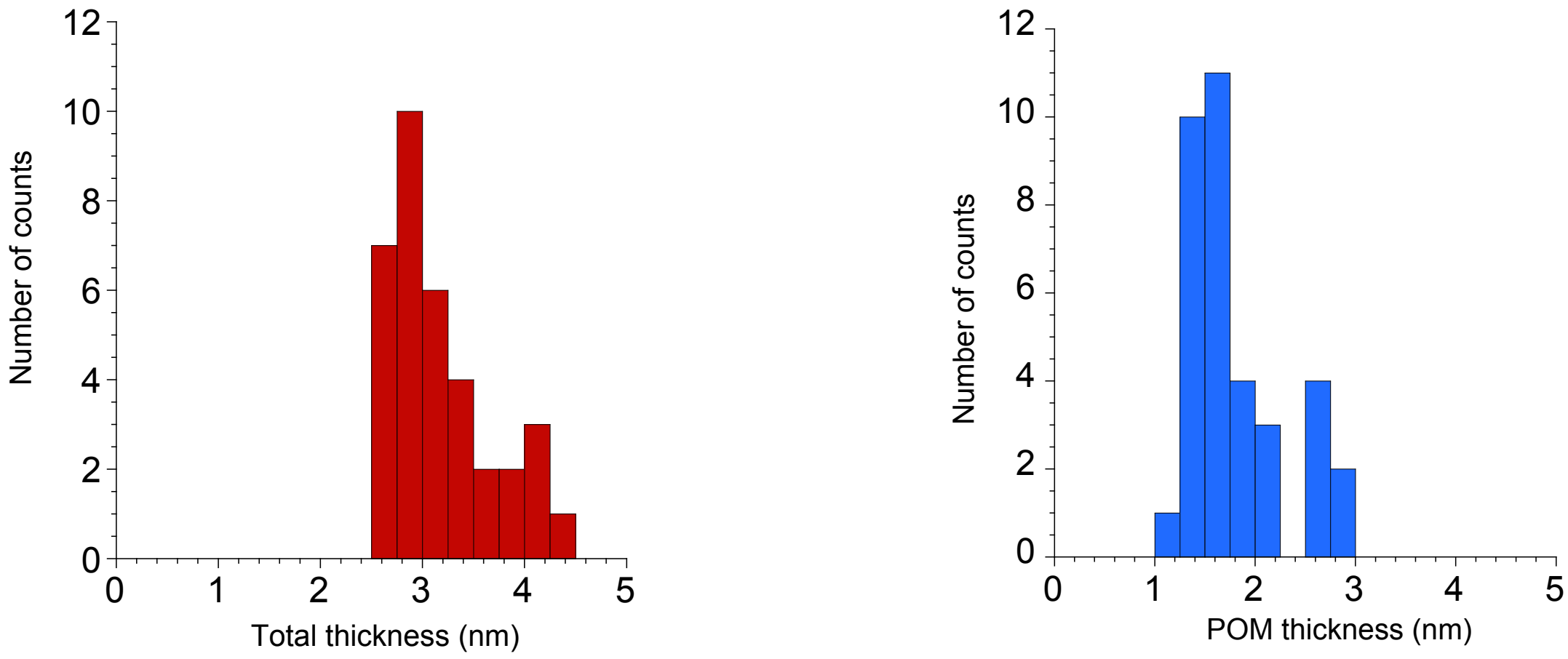

**Figure S12.** Thickness histograms for $K^{W}_{Si}$[COOH]. Left: the total thickness as measured ($SiO_2$ layer plus POM layer); right: the mean $SiO_2$ thickness (1.4 nm) was subtracted to the total thickness values. Most of the $K^{W}_{Si}$[COOH] thickness values range from around 1.5 to 2.5 nm.

## 4. AFM measurements

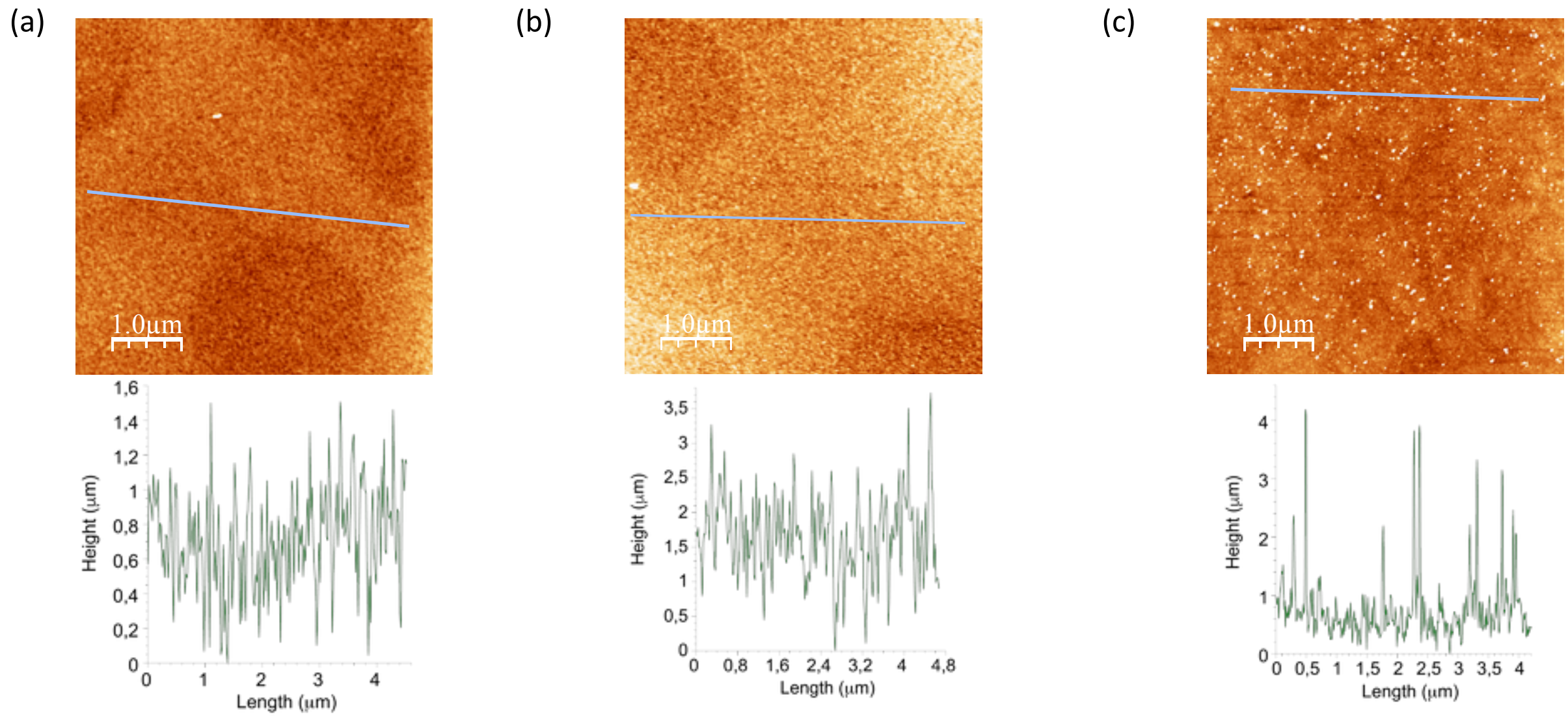

**Figure S13.** 5*5 µm AFM images and Z-profiles of (a) the bare Si/$SiO_2$ substrate, (b) the $K^{W}_{Sn}$[COOH] layer and (c) the $K^{W}_{Si}$[COOH] layer.

## 5. XPS measurements

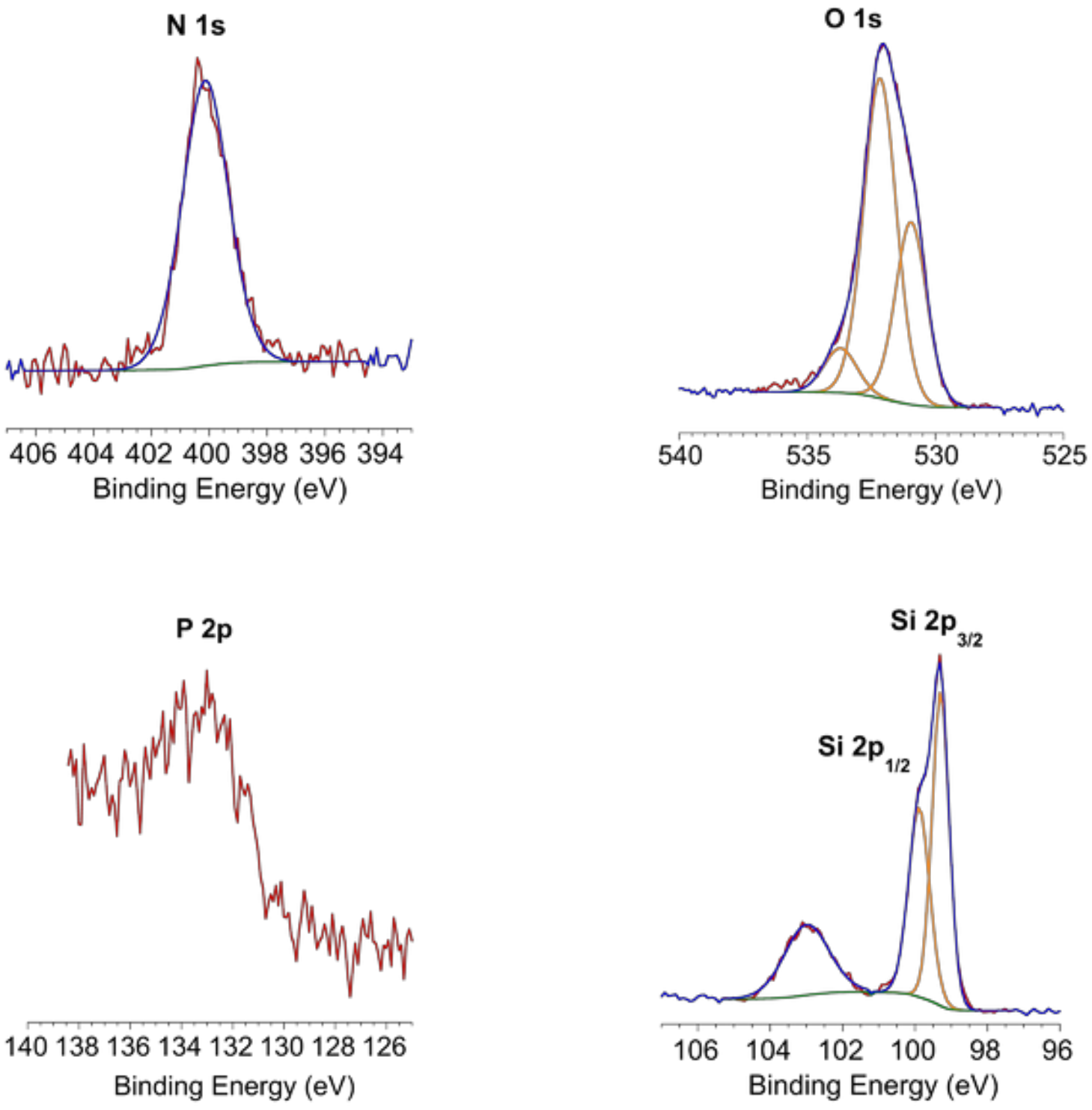

**Figure S14.** N 1s, O 1s, Si 2p and P 2p high-resolution XPS spectra of the $K^{W}_{Sn}$[COOH] layer on $Si/SiO_2$.

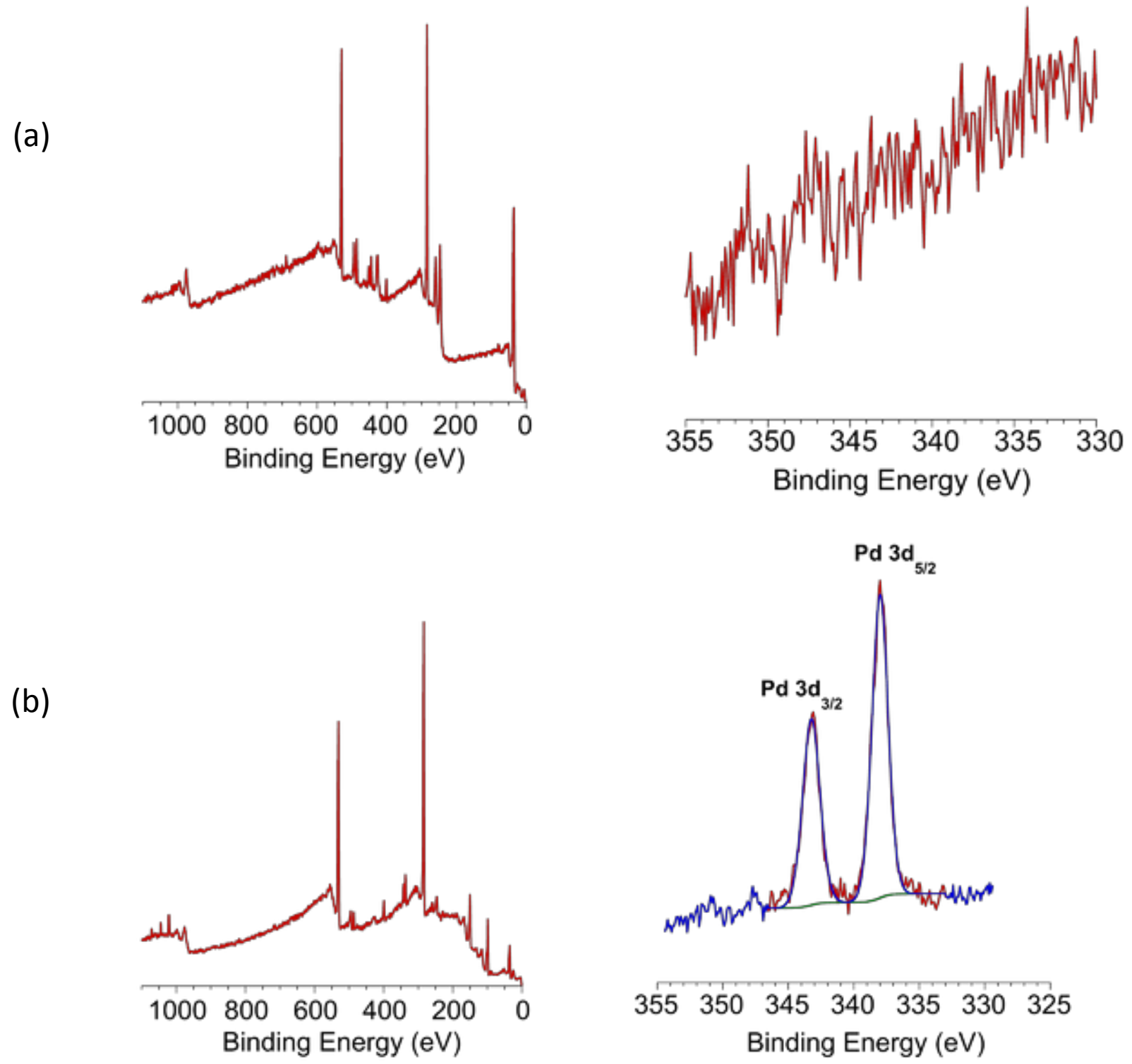

**Figure S15.** Survey and Pd 3d high resolution spectra of the (a) $K^{W}_{Sn}$[COOH] powder and the (b) $K^{W}_{Sn}$[COOH] layer on $Si/SiO_2$.

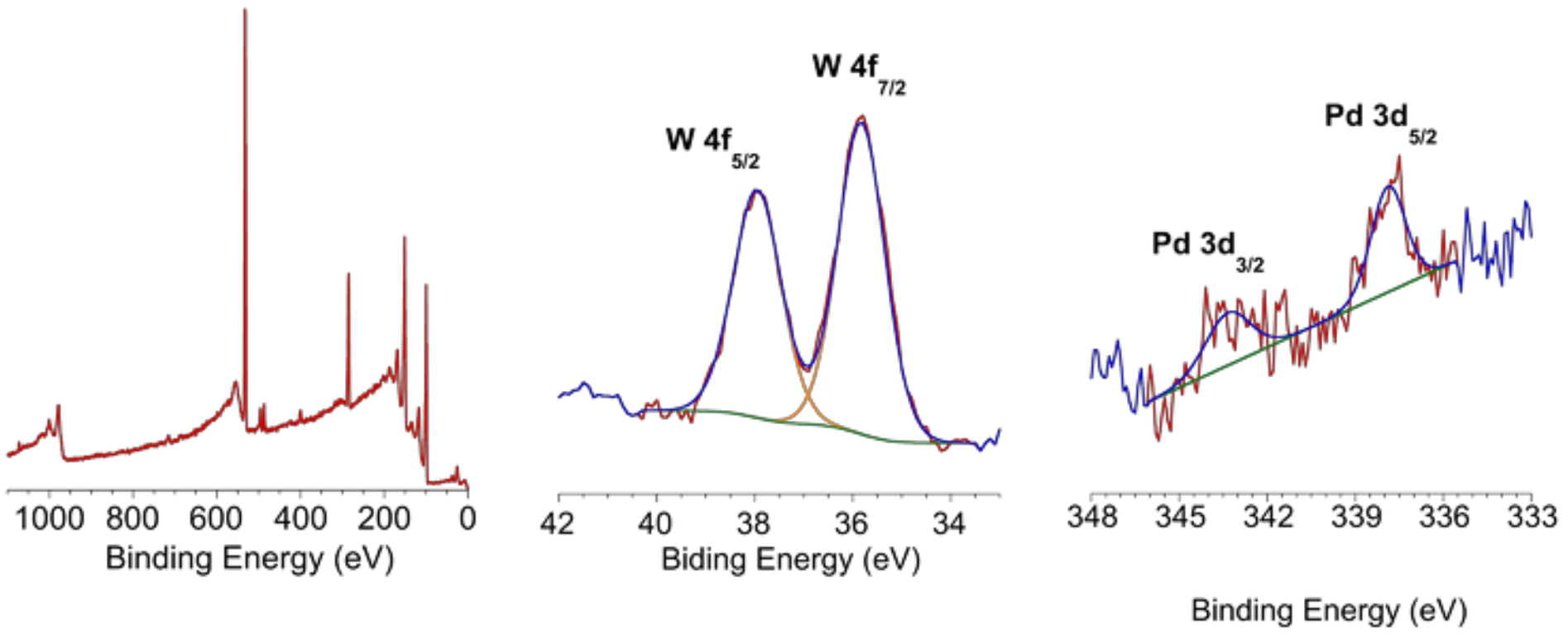

**Figure S16.** Survey and W 4f, Pd 3d high-resolution XPS spectra of the $K^{W}_{Sn}$[H] layer on $Si/SiO_2$.

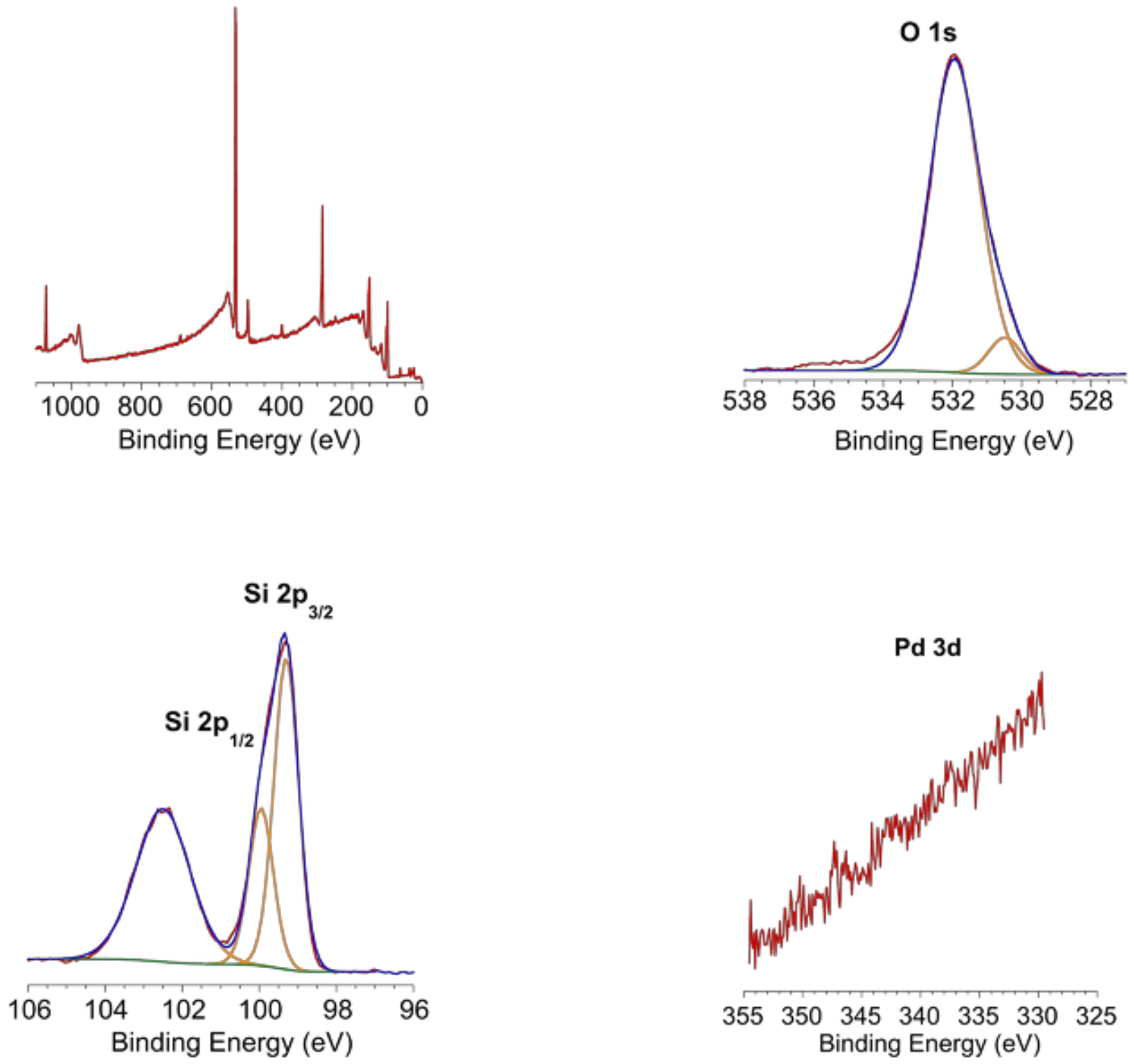

**Figure S17.** Survey, O 1s, Si 2p and Pd 3d high-resolution XPS spectra of the $K^{W}_{Si}$[COOH] layer on $Si/SiO_2$.

## 6. Fits of the modified Simmons model.

The two figures show typical fits of the modified Simmons model on IV of the $K^{W}_{Si}$[COOH] and $K^{W}_{Sn}$[COOH] molecular junctions.

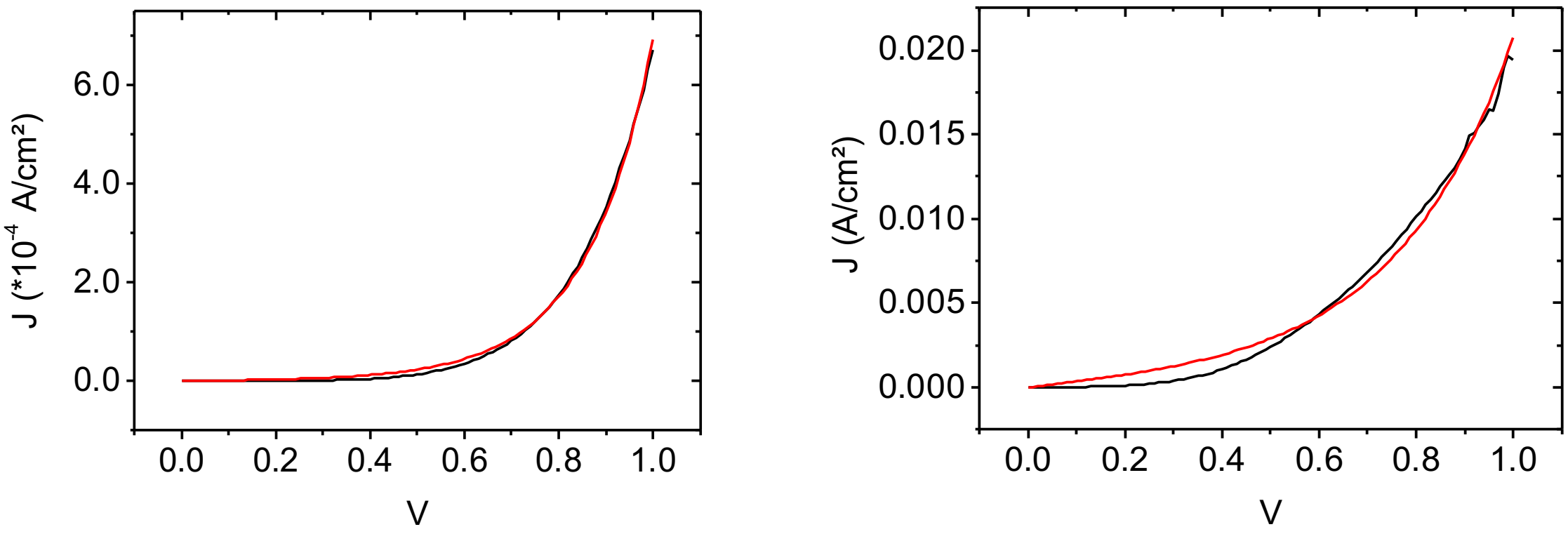

**Figure S18.** (left) Fit (red line) on a IV curve for $K^{W}_{Si}$[COOH] molecular junctions, with $\Phi$ = 1.3 eV and $\alpha$ = 0.82; (right) Fit (red line) on a IV curve for $K^{W}_{Sn}$[COOH] molecular junctions, with $\Phi$ = 1.7 eV and $\alpha$ = 0.37.

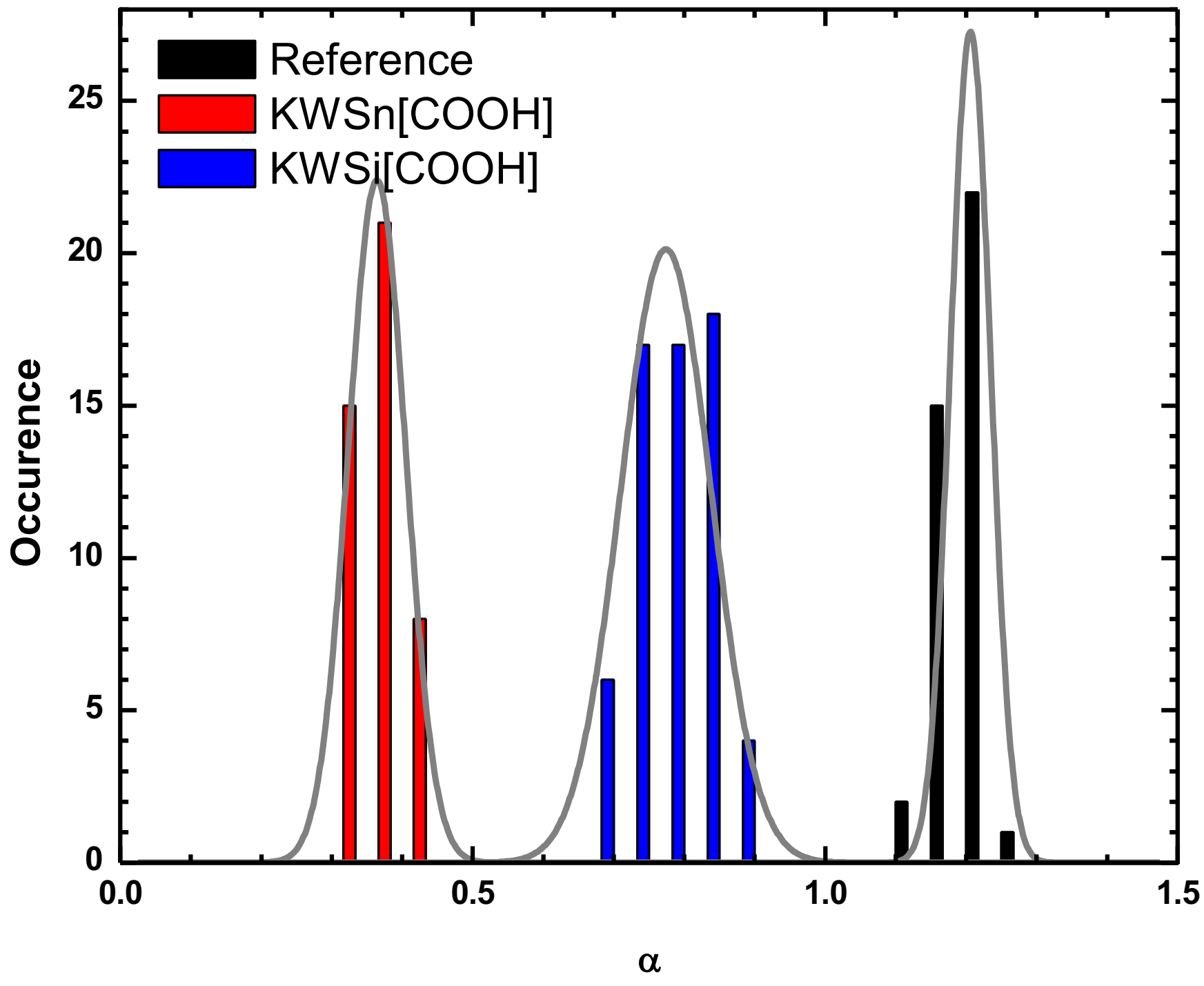

**Figure S19.** α parameter histograms for $K^{W}_{Sn}$[COOH], $K^{W}_{Si}$[COOH] monolayers and the reference bare $Si/SiO_2$ sample. The mean of α parameter α = 1.21 ± 0.06, 0.77 ± 0.13 and 0.36 ± 0.08 and for the $Si/SiO_2$ reference electrode, the $K^{W}_{Si}$[COOH] and the $K^{W}_{Sn}$[COOH] POM junctions, respectively.

## 7. Two layers staircase energy model

We consider a staircase diagram for the energetics of the junction as depicted in figure 12 (right), where the $SiO_2$ barrier is larger than the LUMO of the POMs. In such a case, the Simmons model considers only a simple rectangular energy tunnel barrier with an effective tunnel barrier Φ located between the oxide barrier height and the POM LUMO, $\varepsilon_{POM}$. In the WKB (Wentzel-Kramers-Brilloin) approximation, the effective tunnel barrier Φ for such a staircase tunnel barrier (Figure 12, right) is given by:

$$\sqrt{\Phi}\left(d_{ox}+d_{POM}\right)=\sqrt{\Phi_{ox}}d_{ox}+\sqrt{\varepsilon_{POM}}d_{POM}$$

where $d_{ox}$ and $d_{POM}$ are the thicknesses of the native $SiO_2$ layer and POM monolayers as measured by ellipsometry and $\Phi_{ox}$ (($\Phi_{ox}$ ≈ 1.9 eV, $d_{ox}$ ≈ 1.4 nm) and Φ the values fitted (Figure 11) on the I-V curves for the $Si/SiO_2$ reference sample and the POM samples, respectively.

## 8. XYZ coordinates from DFT calculations

$K^{W}_{Si}$[COOH]

```
74
Energy = -5201.400804792
P    -0.2714791   -2.5173514   -1.1276752
O    -1.0320368   -1.6931649   -0.0620277
O     0.9282867   -3.2857548   -0.4773524
O    -1.2324612   -3.5758760   -1.7834643
O     0.2752175   -1.5880885   -2.2645545
W    -2.9171376   -0.2167145   -0.4426380
W     2.4721859   -4.9013246   -1.7230447
W    -2.4758287   -5.3291261   -0.5053493
W     1.4943006   -2.3962717   -4.3646198
W     2.9197920   -2.3130732    0.6242873
W    -3.3592917   -2.9903453   -2.9620911
W    -2.0166249   -2.5340475    1.9893876
W    -1.0859424   -0.0949023   -3.7385917
W    -0.7825910   -5.2719034   -3.5587788
W     0.7764046   -4.9828712    1.3534474
W     1.9599622    0.2119910   -2.0463179
O    -2.2859565   -1.6400670   -3.8065299
O     0.1725869   -3.8045055   -4.3852842
O    -0.4398735   -3.7123123    2.0953836
O     0.8451607   -5.5422389   -2.5558874
O     0.8899820    0.8553030   -0.5683281
O    -1.9789066    0.3008145   -2.1052120
O    -0.7690516   -5.6117678    0.3486724
O    -3.5123242   -1.7075334   -1.4959069
O     2.4152886   -3.5417357   -3.0955839
O     2.7846755   -1.0153989   -0.8053239
O    -2.6784279   -3.8805068    0.7853093
O     1.6931846   -1.2092997    1.6338386
O    -0.8315390   -1.1890353    2.7087475
O     2.2729410   -3.8005846    1.7119900
O    -2.4485875   -4.3577185   -3.9888926
O     0.4939831    0.8622924   -3.1796057
O     1.9026727   -5.7198094   -0.0719192
O    -1.7690957   -6.1368961   -2.1231705
O    -1.6949324    1.0207759    0.3984319
O     2.4506560   -0.8971063   -3.5475584
O    -3.7278436   -4.3871716   -1.6525201
O    -3.1769344   -1.1900307    1.2100157
O     3.5137623   -3.6585498   -0.6347297
O     0.1375968   -1.1365118   -4.8771390
O    -4.3393682    0.7577797   -0.4726340
O     3.6670968   -5.9997852   -2.2989454
O    -3.4562955   -6.5872059    0.1471011
O     2.3600031   -2.6420545   -5.8335316
O     4.4245503   -1.9263503    1.3700970
O    -4.8734604   -2.8494661   -3.7725625
```

| | | | |
|---|---|---|---|
| O | -2.9063044 | -2.9419478 | 3.4094459 |
| O | -1.7905501 | 1.0023821 | -4.8654642 |
| O | -0.7154528 | -6.5074066 | -4.7570616 |
| O | 0.9424598 | -6.1640611 | 2.5972464 |
| O | 3.1352211 | 1.4612620 | -2.2044957 |
| Si | 0.6365056 | -0.4319494 | 2.6328013 |
| Si | -0.1287627 | 1.5438180 | 0.5314390 |
| O | 0.4064568 | 1.1180612 | 2.0524632 |
| C | -0.1045228 | 3.4088036 | 0.3261199 |
| C | 1.3691520 | -0.3296906 | 4.3618444 |
| C | 0.3711288 | 0.0555751 | 5.4827954 |
| C | -1.1913365 | 4.1800910 | 1.1134632 |
| C | -1.1674842 | 3.8178016 | 2.5786540 |
| C | -0.2957544 | 1.3635928 | 5.1363784 |
| O | -2.1021085 | 3.2537281 | 3.1389884 |
| O | 0.3052612 | 2.4370708 | 5.1852340 |
| O | -1.5308521 | 1.2357351 | 4.6978071 |
| O | -0.0228573 | 4.0907907 | 3.1746896 |
| H | 0.0443637 | 3.5932848 | 4.0556007 |
| H | -1.8490085 | 2.0856587 | 4.2476734 |
| H | -1.0172259 | 5.2674774 | 1.0148670 |
| H | -2.1941091 | 3.9421951 | 0.7282205 |
| H | -0.2318807 | 3.6068899 | -0.7543884 |
| H | 0.9016321 | 3.7699133 | 0.6047904 |
| H | 0.9113453 | 0.1845291 | 6.4366010 |
| H | 2.2014028 | 0.3970978 | 4.3343067 |
| H | 1.8063826 | -1.3168171 | 4.5981504 |
| H | -0.3963980 | -0.7251060 | 5.5999272 |

$K^{W}_{Sn}$[H]

75

Energy = -4764.909502589

| | | | |
|---|---|---|---|
| P | -0.2938615 | -2.5513373 | -1.1725347 |
| O | -1.1641769 | -1.7205715 | -0.1523961 |
| O | 0.8635973 | -3.2432986 | -0.3935841 |
| O | -1.1739161 | -3.6271386 | -1.8720878 |
| O | 0.2942377 | -1.5774140 | -2.2359320 |
| W | -3.2899305 | -0.4269400 | -0.7711303 |
| W | 2.6624515 | -4.6963141 | -1.4154766 |
| W | -2.3935030 | -5.5107032 | -0.7733134 |
| W | 1.8034922 | -2.2327770 | -4.1580374 |
| W | 2.6342599 | -2.0727641 | 0.9549429 |
| W | -3.1654376 | -3.2238945 | -3.3274235 |
| W | -2.4695581 | -2.7203163 | 1.7950053 |
| W | -1.0242645 | -0.1705182 | -3.8162529 |
| W | -0.3362065 | -5.2990697 | -3.6253251 |
| Sn | -0.4776480 | 0.0189223 | 1.1841015 |
| W | 0.6066259 | -4.9229684 | 1.4259136 |
| W | 1.8054473 | 0.3770707 | -1.7739822 |
| O | -2.0834486 | -1.7982246 | -4.0502390 |
| O | 0.5941635 | -3.7537986 | -4.3237152 |
| O | -0.8142941 | -3.7469190 | 2.0256489 |
| O | 1.1806661 | -5.4515823 | -2.4410429 |
| O | 0.6100582 | 0.7699952 | -0.3918679 |
| O | -2.1610568 | 0.1328951 | -2.2978348 |
| O | -0.7614052 | -5.6733124 | 0.2640825 |
| O | -3.6089263 | -1.9849355 | -1.9364679 |
| O | 2.6575339 | -3.3447178 | -2.8049308 |
| O | 2.6923284 | -0.8671885 | -0.5764817 |
| O | -2.8649511 | -4.1473471 | 0.4849953 |
| O | 1.2251663 | -1.0527508 | 1.6289231 |
| O | -1.6142558 | -1.1739187 | 2.4440474 |
| O | 1.9429615 | -3.6282040 | 1.9504214 |
| O | -1.9990477 | -4.5032202 | -4.2330498 |
| O | 0.3975821 | 0.8933230 | -3.0637430 |
| O | 1.9312484 | -5.5864211 | 0.1508279 |
| O | -1.4056435 | -6.2297586 | -2.3033946 |
| O | -2.2652839 | 0.6749998 | 0.3407222 |
| O | 2.5290916 | -0.6813575 | -3.2820963 |
| O | -3.5607126 | -4.6866597 | -2.0914990 |
| O | -3.6703457 | -1.5961740 | 0.7382792 |
| O | 3.4719920 | -3.4395190 | -0.2158066 |
| O | 0.3888591 | -1.0981905 | -4.8233814 |
| O | -4.7511643 | 0.4843775 | -0.9541529 |
| O | 3.9959583 | -5.7129599 | -1.8441072 |
| O | -3.3267320 | -6.8792652 | -0.2700218 |
| O | 2.8378744 | -2.4003725 | -5.5359716 |
| O | 3.9872942 | -1.5830574 | 1.9201828 |
| O | -4.5671469 | -3.2198486 | -4.3430289 |
| O | -3.4456849 | -3.1783201 | 3.1490853 |
| O | -1.6746391 | 0.8983892 | -5.0117767 |
| O | -0.0532676 | -6.5229516 | -4.8151776 |
| C | -0.1926420 | 1.7366903 | 2.4808909 |
| O | 0.7189090 | -6.0893261 | 2.6998940 |
| O | 2.8488368 | 1.7620384 | -1.7825869 |
| C | -0.9281071 | 1.8818554 | 3.6709527 |
| C | 0.6859100 | 2.7655339 | 2.0931645 |
| C | -0.8074595 | 3.0315342 | 4.4510957 |
| C | 0.8191363 | 3.9182581 | 2.8658107 |
| C | 0.0663556 | 4.0737313 | 4.0555989 |
| H | -1.6101978 | 1.0800475 | 3.9740253 |
| H | 1.2566641 | 2.6567253 | 1.1643736 |
| H | 1.4974650 | 4.7195640 | 2.5547906 |
| H | -1.3922591 | 3.1455348 | 5.3702991 |
| C | 0.1734231 | 5.2665474 | 4.8221659 |
| C | 0.2640428 | 6.3051188 | 5.4650729 |
| C | 0.3645489 | 7.5232574 | 6.1875901 |
| C | 1.2608834 | 8.5435467 | 5.7790405 |
| C | 1.3523481 | 9.7437577 | 6.4840620 |
| C | 0.5581829 | 9.9696031 | 7.6170687 |
| C | -0.4301795 | 7.7672851 | 7.3360805 |
| C | -0.3320754 | 8.9708413 | 8.0351214 |
| H | 1.8813279 | 8.3722911 | 4.8948150 |
| H | 2.0528302 | 10.5153849 | 6.1439201 |
| H | 0.6316189 | 10.9141266 | 8.1675811 |
| H | -1.1281390 | 6.9916094 | 7.6640495 |
| H | -0.9611272 | 9.1328169 | 8.9182526 |